\documentclass[galley]{agujournal2018}
\usepackage{apacite}
\usepackage{natbib}
\usepackage{amsmath,amssymb}
\usepackage{subfig,tikz}
\usepackage[normalem]{ulem} 
\usepackage{url} 

\def\dashdotted{\xleaders\hbox to 1em{$- \cdot$}\hfill $-$}

\draftfalse

\journalname{JGR-Planets}

\begin{document}

%
%


\title{Magnetic Induction in Convecting Galilean Oceans}

%
%




\authors{S.~D.~Vance\affil{1}, B.~G.~Bills\affil{1}, C.~J.~Cochrane\affil{1}, K.~M.~Soderlund\affil{2}, N.~G{\'o}mez-P{\'e}rez\affil{3}, M.~J.~Styczinski\affil{4}, and C.~Paty\affil{5}}


\affiliation{1}{Jet Propulsion Laboratory, California Institute of Technology, Pasadena, USA}
\affiliation{2}{Institute for Geophysics, John A. \& Katherine G. Jackson School of Geosciences, The University of Texas at Austin, USA}
\affiliation{3}{British Geological Survey, Edinburgh, UK}
\affiliation{4}{Dept. of Physics, University of Washington, Seattle, USA}
\affiliation{5}{Dept. of Earth Sciences, University of Oregon, Eugene, USA}





\correspondingauthor{S. D. Vance}{svance@jpl.caltech.edu}




\begin{keypoints}
\item Diffusive induction accounting for adiabatic ocean temperatures is distinct in phase and amplitude from induction based on electrical conductivity at the ice-ocean interface 
\item  Based on turbulent global convection models, oceanic flows may generate induced magnetic fields observable by planned spacecraft missions
\item Determining ocean composition from magnetic induction requires additional thermodynamic and electrical conductivity data
\end{keypoints}

%
%


\begin{abstract}
To date, analyses of magnetic induction in putative oceans in Jupiter's large icy moons have assumed uniform conductivity in the modeled oceans. However, the phase and amplitude response of the induced fields will be influenced by the increasing electrical conductivity along oceans' convective adiabatic temperature profiles. Here, we examine the amplitudes and phase lags for magnetic diffusion in modeled oceans of Europa, Ganymede, and Callisto. We restrict our analysis to spherically symmetric configurations, treating interior structures based on self-consistent thermodynamics, accounting for variations in electrical conductivity with depth in convective oceans \citep{Vance_2018}.  The numerical approach considers tens of radial layers.   The induction response of the adiabatic conductivity profile differs from that of an ocean with uniform conductivity set to that at the ice-ocean interface, or to the mean value of the adiabatic profile, by more than 10\% in many cases.  We compare these modeled signals with magnetic fields induced  by oceanic fluid motions that might be used to measure oceanic flows \citep[e.g.,][]{Chave83,Tyler11,Minami17}. For turbulent convection \citep{soderlund2014ocean}, we find that these signals can   dominate induction signal at low latitudes, underscoring the need for spatial coverage in magnetic investigations. Based on end-member ocean compositions \citep{zolotov2008oceanic,zolotov2009chemical}, we quantify the residual magnetic induction signals that might be used to infer the oxidation state of Europa's ocean and to investigate stable liquids under high-pressure ices in Ganymede and Callisto. Fully exploring this parameter space for the sake of planned missions requires electrical conductivity measurements in fluids at low temperature and to high salinity and pressure.
\end{abstract}

%
%

%


%
%
%
%

\section{Introduction}

The jovian system is of particular interest for studying magnetic induction in icy ocean worlds. Jupiter has a strong magnetic field whose dipole axis is tilted $9.5^\circ$ with respect to its rotation axis \citep{Acuna_1976}, while the orbits of the Galilean moons lie very nearly in the equatorial plane of
Jupiter. This means that Jupiter?s magnetic field varies in time at the orbital positions of the satellites. Also, the outer layers of the satellites themselves are believed to consist mainly of water ice at the surface, underlain by salty oceans. Brines are good conductors, while ice is a significant insulator.

Magnetic induction from Jupiter's diurnal signal sensed by the  \textit{Galileo} mission provides the most compelling direct observational evidence for the existence of oceans within Europa and Ganymede \citep{Saur_1998,khurana1998induced,kivelson2000galileo,schilling2007time,hand2007empirical,khurana2009electromagnetic}. The case has also been made for an induction response from an ocean in Callisto \citep{zimmer2000subsurface}, but this interpretation is clouded by possible ionospheric interference \citep{Liuzzo_2015, Hartkorn_2017}.

Longer period signals penetrate more deeply, as penetration of the magnetic field into the interior is a diffusive process.  It is convenient that the skin depths at the dominant periods of variation experienced by Europa, Ganymede, and Callisto are comparable to the expected ocean depths, which makes it possible to probe the properties of their oceans using magnetic induction. The spectrum of frequencies driving induced magnetic responses includes not just the orbits of the Galilean satellites and the rotation of Jupiter's tilted dipole field, but also their harmonics and natural oscillations \citep{Saur_2009,seufert2011multi}. Electrical conductivity structure within the subsurface oceans---for example, from convective adiabatic temperature gradients \citep{Vance_2018} and stratification \citep{vance2009oceanography}---will respond at these frequencies. 

Further variations in the magnetic fields arise from the motion of the moons about Jupiter.  Perturbations to the orbits of the moons arise from multiple sources, including the oblate figure of Jupiter, gravitational interactions with the other satellites, and even from Saturn and the Sun \citep{lieske1998galilean,lainey2006srg}.

Here, we examine the amplitudes and phase lags for magnetic diffusion in modeled oceans of Europa, Ganymede, and Callisto. We restrict our analysis to spherically symmetric configurations, treating interior structures based on self-consistent thermodynamics, which account for variations in electrical conductivity with depth in convective oceans \citep{Vance_2018}. 
In addition, we consider the generation of induced magnetic fields by oceanic fluid motions that may bias the interpretation of a satellite's magnetic behavior if not accommodated and which, more optimistically, might be used to probe the ocean flows directly \citep[e.g.,][]{Chave83,Tyler11,Minami17}. Based on end-member ocean compositions \citep{zolotov2008oceanic,zolotov2009chemical}, we demonstrate the possibilities for using magnetic induction to infer the oxidation state of Europa's ocean and to identify stable liquid layers under high-pressure ices in Ganymede and Callisto.

In Section~\ref{section:inductionDerivation} we describe a numerical method for computing the induction response.  Section~\ref{section:diffusive} examines the diffusive induction response of Jupiter's ocean moons, first describing the  frequency content of temporal variations in Jupiter's field in the reference frames of the Galilean moons (S~\ref{subsection:frequencies}), then the  interior structure models that include layered electrical conductivity consistent with the modeled compositions (S~\ref{subsection:interiors}). In Section~\ref{subsection:amplitudes}, we detail the corresponding amplitude and phase responses of the diffusive magnetic induction, and finally in Section~\ref{subsection:residuals}, we compare the diffusive fields to the field imposed by Jupiter. Section~\ref{section:flows} describes simulations of oceanic flows (S~\ref{subsection:motions}) and resulting magnetic induction (S~\ref{subsection:flowinduction}) that  adds to the diffusive component. Section~\ref{section:discussion} describes the prospects for detecting these different signals.

\section{Induction Response Model}\label{section:inductionDerivation}

We are interested in the magnetic fields induced within a spherically
symmetric body, in which electrical conductivity is a piece-wise constant
function of distance from the center. We thus assume bounding radii%
\begin{equation}
\{r_{1},r_{2},r_{3},\cdots ,r_{m}\}
\end{equation}%
where 
\begin{equation}
r_{m}=R
\end{equation}%
is the outer radius of the spherical body.

The corresponding conductivity values are%
\begin{equation}
\{\sigma _{1},\sigma _{2},\sigma _{3},\cdots ,\sigma _{m}\}
\end{equation}

We also assume that there is an imposed external magnetic potential,
represented by a sum of terms, each of which has the form%
\begin{equation}\label{equation:magpotential}
\Phi[r,\theta ,\phi ,t]=R\ B_{e}\left( \frac{r}{R}\right)
^{n}S_{n,m}[\theta ,\phi ]\ \exp [-i\ \omega \ t]
\end{equation}%
where $\{r,\theta ,\phi \}$ are spherical coordinates ($r$ is radius, $%
\theta $ is colatitude, and $\phi $ is longitude) of the field point, $B_{e}$
is a scale factor, $S_{n,m}[\theta ,\phi ]$ is a surface spherical harmonic
function of degree $n$ and order $m$, while $t$ is time and $\omega $ is the
frequency of oscillation of the imposed potential.

Within each layer, the magnetic field vector $B$ must satisfy the
differential equation%
\begin{equation}
\nabla ^{2}B=-k^{2}B  \label{diffusion}
\end{equation}%
where $k$ is a scalar wavenumber given by%
\begin{equation}\label{equation:k}
k^{2}=i\ \omega \ \mu_0 \ \sigma
\end{equation}
where $\omega $ is frequency, $\sigma $ is electrical conductivity, and the
magnetic constant (permeability of free space) is given by%
\begin{equation}
\mu_0 =4\pi \times 10^{-7} N/A^{2}
\end{equation}%
 with units $N$ and $A$ being Newton and Ampere.

\subsection{Radial Basis Functions}

The poloidal component of the magnetic field inside the body is given by
sums of terms with the forms%
\begin{equation}
B_{r}[r,\theta ,\phi ,t]=\frac{C}{r}\left( F[r]\right) \ n(n+1)\
S_{n,m}[\theta ,\phi ]\ \exp [-i\ \omega \ t]  \label{rad comp interior}
\end{equation}%
\begin{equation}
B_{\theta }[r,\theta ,\phi ,t]=\frac{C}{r}\left( \frac{d\ rF[r]}{dr}\right)
\ \ \frac{dS_{n,m}[\theta ,\phi ]\ }{d\theta }\exp [-i\ \omega \ t]
\label{th comp interior}
\end{equation}%
\begin{equation}
B_{\phi }[r,\theta ,\phi ,t]=\frac{C}{r\sin [\theta ]}\left( \frac{d\ rF[r]}{%
dr}\right) \ \ \frac{dS_{n,m}[\theta ,\phi ]\ }{d\phi }\exp [-i\ \omega \ t]
\label{ph comp interior}
\end{equation}%
where $C$ is a constant, and $F[r]$ is a function of radius, which we need
to determine.

Applying separation of variables to the governing differential equation (\ref%
{diffusion}), one finds that the radial factor $F[r]$ in the solution must
satisfy the ordinary differential equation 
\begin{equation}
\frac{d^{2}F}{dr^{2}}+\left( \frac{2}{r}\right) \frac{dF}{dr}+(k^{2}-\frac{%
n(n+1)}{r^{2}})F=0
\end{equation}%
This is a second order equation having two solutions:
\begin{eqnarray}
F_{n}^{+}[r]=j_{n}[k\ r]\\
F_{n}^{-}[r]=y_{n}[k\ r]
\end{eqnarray}%
where $j_{n}[x]$ is a spherical Bessel function of the first kind of order $%
n$, and argument $x$, and $y_{n}[x]$ is a spherical Bessel function of the second kind.

It will also be convenient to define another set of related functions%
\begin{eqnarray}
G_{n}^{+}[r] &=&\frac{d}{dr}\left( r\ F_{n}^{+}[r]\right) \\
&=&(n+1)\ \ j_{n}[k\ r]-\left( k\ r\right) \ j_{n+1}[k\ r]  \nonumber
\end{eqnarray}%
and%
\begin{eqnarray}
G_{n}^{-}[r] &=&\frac{d}{dr}\left( r\ F_{n}^{-}[r]\right) \\
&=&(n+1)\ \ y_{n}[k\ r]-\left( k\ r\right) \ y_{n+1}[k\ r]  \nonumber
\end{eqnarray}

In the magnetic induction problem, as applied to the Galilean satellites, the only case of
interest is for an imposed dipole field, where $n=1$. In that case, the
radial basis functions for the radial component of the field, are%
\begin{eqnarray}
F_{1}^{+}[k\ r] &=&j_{1}[k\ r] \\
&=&\frac{\sin [k\ r]-\left( k\ r\right) \cos [k\ r]}{\left( k\ r\right) ^{2}}
\nonumber
\end{eqnarray}%
and%
\begin{eqnarray}
F_{1}^{-}[k\ r] &=&y_{1}[k\ r] \\
&=&\frac{-\cos [k\ r]-\left( k\ r\right) \sin [k\ r]}{\left( k\ r\right) ^{2}%
}  \nonumber
\end{eqnarray}%
In similar fashion, the radial basis functions for the transverse components
are%
\begin{eqnarray}
G_{1}^{+}[k\ r] &=&2\ j_{1}[k\ r]-(k\ r)\ j_{2}[k\ r] \\
&=&\frac{\left( k\ r\right) \cos [k\ r]-\left( 1-k^{2}\ r^{2}\right) \sin
[k\ r]}{\left( k\ r\right) ^{2}}  \nonumber
\end{eqnarray}%
and%
\begin{eqnarray}
G_{1}^{-}[k\ r] &=&2\ y_{1}[k\ r]-(k\ r)\ y_{2}[k\ r] \\
&=&\frac{\left( k\ r\right) \sin [k\ r]+\left( 1-k^{2}\ r^{2}\right) \cos
[k\ r]}{\left( k\ r\right) ^{2}}  \nonumber
\end{eqnarray}

In both cases, the latter form is singular at the origin ($r=0$), so in the
innermost spherical layer, we only use $F^{+}[k\ r]$ and $G^{+}[k\ r].$ In
other layers, we use linear combinations of $F^{+}$ and $F^{-}$ and linear
combinations of $G^{+}$ and $G^{-}$.

\subsection{Internal Boundary Conditions}

The resulting piece-wise-defined radial functions characterize the
radial part of the magnetic field. The radial component has the form%
\begin{equation}
F[r]=\left\{ 
\begin{array}{cccc}
c_{1}\ F^{+}[k_{1}r] &  & \mathrm{if} & 0<r\leq r_{1} \\ 
c_{2\ }F^{+}[k_{2}r] & +\ d_{2}\ F^{-}[k_{2}r] & \mathrm{if} & r_{1}<r\leq r_{2} \\ 
c_{3}\ F^{+}[k_{3}r] & +\ d_{3}\ F^{-}[k_{3}r] & \mathrm{if} & r_{2}<r\leq r_{3} \\ 
&  &  &  \\ 
c_{m}\ F^{+}[k_{m}r] & +\ d_{m}\ F^{-}[k_{m}r] & \mathrm{if} & r_{m-1}<r\leq r_{m}%
\end{array}%
\right.  \label{pw radial}
\end{equation}%
The transverse components yield similar structure, but with $G$ replacing $F$.%

The constants $c_{j}$ and $d_{j}$ are determined by continuity of radial ($r$%
) and transverse ($\theta ,\phi $) components of the magnetic field across
the boundaries. For each internal boundary, it must hold that%
\begin{eqnarray}
F[r_{j}] &=&c_{j}\ F^{+}\left[ k_{j}\ r_{j}\right] +d_{j}\ F^{-}\left[
k_{j}\ r_{j}\right]  \label{int BC rad} \\
&=&c_{j+1}\ F^{+}\left[ k_{j+1\ }r_{j}\right] +d_{j+1}\ F^{-}\left[ k_{j+1\
}r_{j}\right]  \nonumber
\end{eqnarray}%
to ensure continuity of the radial component of the magnetic field, and%
\  likewise for $G$
to ensure continuity of the transverse components. These continuity
constraints yield two equations at each internal boundary, from which we can
determine the layer coefficients.

The internal boundary conditions are only part of the story. In a model with 
$m$ layers, we have $2m-1$ coefficients to determine (recall that $d_{1}=0,$
to avoid singular behavior at the origin), but only $m-1$ internal
boundaries, and thus only $2m-2$ constraints. The external boundary
condition provides the additional information to make the problem evenly
determined.

Even without the external boundary condition,  a provisional
solution is obtained by setting $c_{1}=1$ and using the internal boundary constraints
to determine the other coefficient values. Using notation similar to that of
Parkinson (1983, page 314), we can write a recursion relation that
transforms the coefficients in the $j$th layer into those for the layer
above it%
\begin{equation}\label{equation:recursionEqn}
\left[ 
\begin{array}{c}
c_{j+1} \\ 
d_{j+1}%
\end{array}%
\right] =T_{j}[k_{j},k_{j+1},r_{j}]\cdot \left[ 
\begin{array}{c}
c_{j} \\ 
d_{j}%
\end{array}%
\right]
\end{equation}%
where the transformation matrix has elements%
\begin{equation}\label{equation:recursionCoeffs}
T_{j}[k_{j},k_{j+1},r_{j}]=\frac{1}{\alpha _{j}}\left[ 
\begin{array}{cc}
\beta _{j} & \gamma _{j} \\ 
\delta _{j} & \varepsilon _{j}%
\end{array}%
\right]
\end{equation}%
with 
\begin{equation}
\alpha _{j}=F^{+}\left[ k_{j+1\ }r_{j}\right] \ast G^{-}\left[ k_{j+1\ }r_{j}%
\right] -F^{-}\left[ k_{j+1\ }r_{j}\right] \ast G^{+}\left[ k_{j+1\ }r_{j}%
\right]
\end{equation}%
which is a function of the conductivity in the layer above the boundary only.
The other elements  depend on the conductivities on both sides of the
boundary%
\begin{equation}
\beta _{j}=F^{+}\left[ k_{j\ \ }r_{j}\right] \ast G^{-}\left[ k_{j+1\ }r_{j}%
\right] -F^{-}\left[ k_{j+1\ }r_{j}\right] \ast G^{+}\left[ k_{j\ \ }r_{j}%
\right]
\end{equation}%
\begin{equation}
\gamma _{j}=F^{-}\left[ k_{j\ \ }r_{j}\right] \ast G^{-}\left[ k_{j+1\ }r_{j}%
\right] -F^{-}\left[ k_{j+1\ }r_{j}\right] \ast G^{-}\left[ k_{j\ \ }r_{j}%
\right]
\end{equation}%
and%
\begin{equation}
\delta _{j}=F^{+}\left[ k_{j+1\ }r_{j}\right] \ast G^{+}\left[ k_{j\ \ }r_{j}%
\right] -F^{+}\left[ k_{j\ }\ r_{j}\right] \ast G^{+}\left[ k_{j+1\ }r_{j}%
\right]
\end{equation}%
\begin{equation}
\varepsilon _{j}=F^{+}\left[ k_{j+1\ }r_{j}\right] \ast G^{-}\left[ k_{j\ \
}r_{j}\right] -F^{-}\left[ k_{j\ \ }r_{j}\right] \ast G^{+}\left[ k_{j+1\
}r_{j}\right]
\end{equation}

We thus start in the central spherical layer, with $c_{1}=1$ and $d_{1}=0$,
and then propagate upward through the stack of layers until we have the
coefficients in each of the $m$ layers. This set of layer coefficients, with
the radial basis functions, yields structures as given in equations (\ref{equation:recursionEqn}) and (\ref{equation:recursionCoeffs}).

\subsection{External Boundary Conditions}

The final step is matching the external surface boundary condition. Outside
the sphere, the magnetic field is represented by a scalar potential which is
the sum of an imposed external contribution and an induced internal
contribution. That sum has spatial dependence given by the form%
\begin{equation}
\Phi[r,\theta ,\phi ]=R\left( B_{e}\left( \frac{r}{R}\right) ^{n}+B_{i}\left( 
\frac{R}{r}\right) ^{n+1}\right) S_{n}[\theta ,\phi ]
\end{equation}%
We have dropped the subscript $m$ from $S_{n,m}$ because a suitable
choice of axes results in $m=0$ for both external and internal fields for
the case of spherical symmetry we consider here.
The vector field is obtained from the potential via%
\begin{equation}
B=-\nabla \Phi
\end{equation}%
The radial component of the vector field, evaluated at the surface ($r=R$),
is%
\begin{equation}
B_{r}=-\left( n\ B_{e}-(n+1)B_{i}\right) S_{n}[\theta ,\phi ]
\end{equation}%
and the tangential components are%
\begin{equation}
B_{\theta }=-(B_{e}+B_{i})\frac{\partial S_{n}[\theta ,\phi ]}{\partial
\theta }
\end{equation}%
and%
\begin{equation}
B_{\phi }=-(B_{e}+B_{i})\frac{1}{\sin [\theta ]}\frac{\partial S_{n}[\theta
,\phi ]}{\partial \phi }
\end{equation}

Matching these with the corresponding interior components, as given in
equations (\ref{rad comp interior}), (\ref{th comp interior}), and (\ref{ph
comp interior}), but evaluated at the top of the upper-most layer, we obtain%
\begin{equation}
-(n\ B_{e}-(n+1)\ B_{i})R=n\ (n+1)\ (c_{m}\ F^{+}[k_{m}\ R]+d_{m}\
F^{-}[k_{m}\ R])
\end{equation}%
and%
\begin{equation}
-(B_{e}+B_{i})R=\left( c_{m}\ G^{+}[k_{m}\ R]+d_{m}\ G^{-}[k_{m}\ R]\right)
\end{equation}

From these two equations, we can first solve for $B_{e}$ and $B_{i}$. The
result is%
\begin{equation}
\widehat{B}_{e}= \frac{-1}{R(2n+1)} \left( c_{m}\ A_{m}+d_{m}\
B_{m}\right)  \label{Be from layers}
\end{equation}%
\begin{equation}
\widehat{B}_{i}= \frac{1}{R(2n+1)} \left( c_{m}\ C_{m}+d_{m}\
D_{m}\right)  \label{Bi from layers}
\end{equation}%
where we introduce $\widehat{B}_{e}$ and $\widehat{B}_{i}$ to distinguish 
solutions in terms of internal properties from the external and induced 
magnetic moments.
We also define the parameters $A_{m}$, $B_{m}$, $C_{m}$, and $D_{m}$ by%
\begin{eqnarray}
A_{m} &=&\left( n+1\right) \left( n\ F^{+}[k_{m}\ R]+\ G^{+}[k_{m}\ R]\right)
\\
B_{m} &=&\left( n+1\right) \left( n\ F^{-}[k_{m}\ R]+\ G^{-}[k_{m}\ R]\right)
\nonumber
\end{eqnarray}%
and%
\begin{eqnarray}
C_{m} &=&n\left( \left( n+1\right) F^{+}[k_{m}\ R]-G^{+}[k_{m}\ R]\right) \\
D_{m} &=&n\left( \left( n+1\right) F^{-}[k_{m}\ R]-G^{-}[k_{m}\ R]\right) 
\nonumber
\end{eqnarray}

As previously noted, choice of $c_{1}=1$ permits solution of layer 
coefficients $c_{j}$ and $d_{j}$ relative to each other with only knowledge 
of the interior properties. We can then solve for $\widehat{B}_{e}$ and
$\widehat{B}_{i}$ in terms of the interior structure quantities $k_{j}$
and $r_{j}$. We can then conveniently relate this to the magnetic field
that will be induced from the conducting body for a given external
field $B_{e}^{\ast }$ by introducing a scale factor:
\begin{equation}
S=\frac{B_{e}^{\ast }}{\widehat{B}_{e}}
\end{equation}%
Choosing a normalized value of 
\begin{equation}
B_{e}^{\ast }=1
\end{equation}%
means that physically correct layer coefficients may be determined by
multiplying the magnitude of the applied external field to the coefficients 
$c_{j}^{\ast }$ and $d_{j}^{\ast }$, obtained from%
\begin{equation}
\left[ 
\begin{array}{c}
c_{j}^{\ast } \\ 
d_{j}^{\ast }%
\end{array}%
\right] =S\ \left[ 
\begin{array}{c}
c_{j} \\ 
d_{j}%
\end{array}%
\right]
\end{equation}%
For an applied external field $B_{e}^{\ast }$ in real units, the physical
magnetic field within each layer is then given by 
\begin{eqnarray}
B_{r,j}[r,\theta ,\phi ,t] &=& \frac{B_{e}^{\ast }}{r}\left( c_{j}^{\ast }F^{+}[k_{j}r]
+ d_{j}^{\ast }F^{-}[k_{j}r] \right) n(n+1)
S_{n}[\theta ,\phi ]\ \exp [-i\ \omega \ t]
\nonumber
\\
B_{\theta,j}[r,\theta ,\phi ,t] &=& \frac{B_{e}^{\ast }}{r}\left( 
c_{j}^{\ast }G^{+}[k_{j}r] + d_{j}^{\ast }G^{-}[k_{j}r] \right)
 \frac{dS_{n}[\theta ,\phi ]\ }{d\theta }\ \exp [-i\ \omega \ t]
\\
B_{\phi,j}[r,\theta ,\phi ,t] &=& \frac{B_{e}^{\ast }}{r\sin [\theta ]}\left( 
c_{j}^{\ast }G^{+}[k_{j}r] + d_{j}^{\ast }G^{-}[k_{j}r] \right) 
 \frac{dS_{n}[\theta ,\phi ]\ }{d\phi }\ \exp [-i\ \omega \ t]
\nonumber
\end{eqnarray}%

The ratio of internal and external field strengths at the exterior surface is
given from equations (\ref{Be from layers}) and (\ref{Bi from layers}) via%
\begin{equation}
Q\equiv \frac{\widehat{B}_{i}}{\widehat{B}_{e}}=-\frac{c_{m}^{\ast }\ 
C_{m}+d_{m}^{\ast }\ D_{m}}{c_{m}^{\ast }\ A_{m}+d_{m}^{\ast }\ B_{m}}
\end{equation}

In \citet{zimmer2000subsurface} and \citet{khurana2009electromagnetic}, this complex ratio is
written as the product of a real magnitude and a phase shift:%
\begin{equation}
Q=A^{\ast}\exp [i\ \gamma^{\ast} ]
\end{equation}%
where $A^{\ast}$ is a positive real number representing amplitude and $\gamma^{\ast} 
$ is a real number representing the phase of the induced field relative
to the imposed field.

 In the aforementioned previous work, an explicit formula is given for the
 result from a 3-layer model, in which the conductivities in the innermost ($%
 j=1$) and outermost ($j=3$) layers are zero, and the middle layer (intended
 to represent a salty ocean in Europa) has a finite conductivity. In this
 model, there are essentially four free parameters---3 bounding radii ($%
 r_{1},r_{2},r_{3}$) and a middle layer conductivity ($\sigma _{2}$)---that
 determine the critical wavenumber ($k_{2}$). We refer to this model as the ocean-only model.

 In our notation, the resulting ratio $Q$ for the ocean-only model is%
\begin{eqnarray}
 Q &=& \frac{-n}{n+1}\ 
   \frac{ j_{n+1}[k_2\ r_1] \ast y_{n+1}[k_2\ r_2] - j_{n+1}[k_2\ r_2] \ast y_{n+1}[k_2\ r_1]}{
          j_{n+1}[k_2\ r_1] \ast y_{n-1}[k_2\ r_2] - j_{n-1}[k_2\ r_2] \ast y_{n+1}[k_2\ r_1] }
%
 \label{three_shell_model}
\end{eqnarray}%
Because we know the complex phase of the wavenumber $k$, we can use properties of Bessel 
 functions to solve for the amplitude and phase for the induced magnetic field.
 We defined $k^2 = i\omega\mu\sigma$ (Eq.~\ref{equation:k}), so $k = \exp [i\pi/4]\sqrt{\omega\mu\sigma}$.
 The (real) magnitude of $k$ is $|k| = \sqrt{\omega\mu\sigma}$, and all layers will have the
 same complex phase $\pi/4$. We can therefore express the wavenumber for each layer as
 \begin{equation}
     k_j = \kappa_j \exp [i\pi/4],\ \ \kappa_j = \sqrt{\omega\mu_j \sigma_j}
 \end{equation}
 When $\kappa_2 r_2$ is large, $j_{n+1}[\kappa_2\, r_2] = -j_{n-1}[\kappa_2\ r_2]$
 and $y_{n+1}[\kappa_2\, r_2] = -y_{n-1}[\kappa_2\ r_2]$. We can make use of these relations to 
 note that the amplitude and phase for the induced magnetic field for a perfectly conducting
 sphere of radius $r_2$ will be $n/(n+1)$ and $0$, respectively. Thus, we can also define 
 an amplitude and phase for the induction response relative to those for a perfectly
 conducting sphere of radius $R$:
 \begin{equation}
 A = A^{\ast}\ \frac{n+1}{n}\, \left(\frac{r_2}{R}\right)^3,\ \ \
 \gamma = \gamma^{\ast}
 \label{relative_amplitude}
\end{equation}
 A perfectly conducting sphere of radius $R$ therefore has a relative amplitude of $A=1$
 and $\gamma=0$.

\section{Diffusive Induction in Jupiter's Ocean Moons} \label{section:diffusive}
\subsection{Spectral Content of the Imposed Magnetic Field Variations}\label{subsection:frequencies}

\begin{figure}[t]
\centering
\includegraphics[width=30pc]{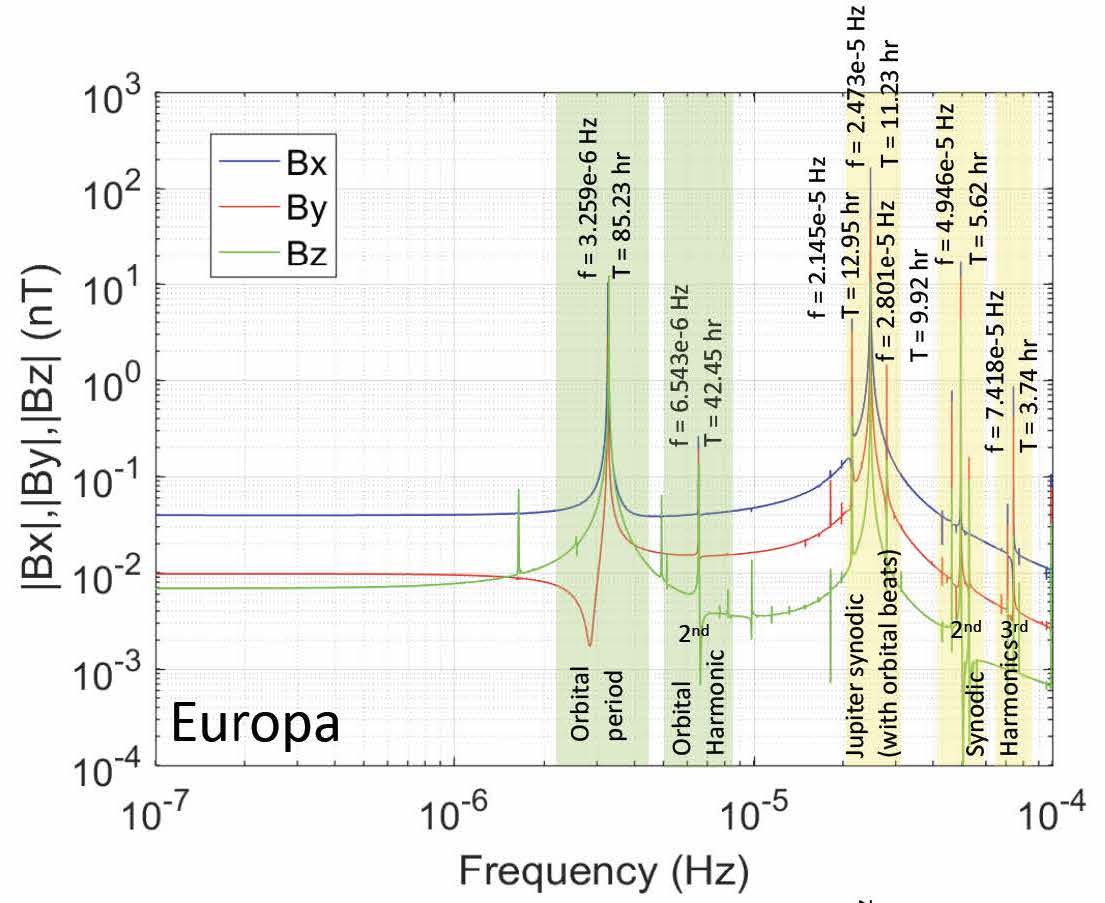}
\caption{Europa: Variations in orbital parameters over time introduce magnetic fluctuations at multiple frequencies beyond the Jupiter rotation and satellite orbital frequencies. The different vector components contain unique information at multiple frequencies resulting from the harmonics and beats of the orbital and rotational oscillations.}
\label{figure:frequencies}
 \end{figure}

Temporal variations in the magnetic field occur in the reference frames of Jupiter's satellites. Figure~\ref{figure:frequencies} shows the strongest components, arising from the orbital and synodic periods and their harmonics. \citet{seufert2011multi} determined the frequency spectra for the time-varying magnetic perturbations applied to each of the four Galilean moons based on the VIP4 model of \citet{connerney1998new} and the Jovian current sheet model of \citet{Khurana_1997}.  \citet{seufert2011multi} also examined the frequency spectra of magnetic perturbations from dynamic migration of the Jovian magnetopause based on solar wind data from the Ulysses spacecraft, which we do not consider here.

To calculate the frequencies, we first compute the magnetic field using the JRM09 Jupiter field model accounting for Juno measurements \citep{Connerney_2018} and using the plasma sheet model from \citet{Khurana_1997}. We then compute the field at the orbital positions of the moons using the most recent and up-to-date NAIF-produced spice kernels and three years of data covering the duration of the Europa Clipper mission (tour 17F12v2). Finally, we compute the Fourier transform of the entire data sets to determine the induction frequencies. 


The temporal variations in imposed magnetic field at each satellite depend on the
orbits of the satellites and the magnetic field of Jupiter. 
To find them, we compute Jupiter's magnetic field
in a Jupiter-centered coordinate system from a spherical harmonic series representation of the magnetic potential, which is a variant of Eq.~\ref{equation:magpotential}:%
\begin{equation}
\Phi \lbrack r,\theta ,\phi ]=R\sum_{n=1}\left( \frac{R}{r}\right)
^{n+1}\sum_{m=0}^{n}P_{n,m}[\sin [\theta ]]\ \left( g_{n,m}\cos [m\phi
]+h_{n,m}\sin [m\phi ]\right).
\end{equation}
The magnetic field vector is the negative
gradient of the scalar potential%
\begin{eqnarray}
B &=&-\nabla \Phi \\
&=&-\left\{ \frac{\partial \Phi }{\partial x},\frac{\partial \Phi }{\partial
y},\frac{\partial \Phi }{\partial z}\right\}
\end{eqnarray}
The mean radius is $R=71,492$~km. The rotation rate of Jupiter, as defined in the System III longitude \citep{seidelmann1977evaluation}, is $\omega =870.536^{\circ }$/day.

 \begin{figure}[t]
\centering
\includegraphics[width=30pc]{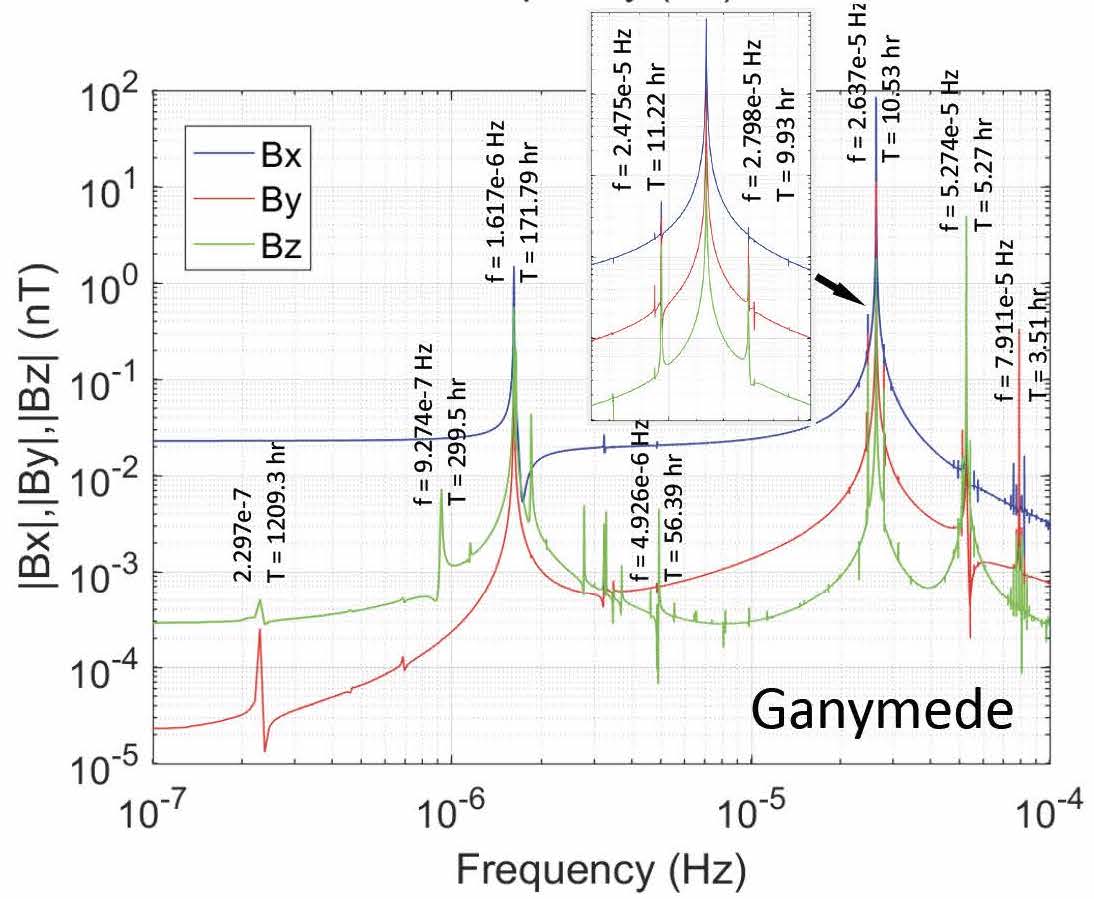}
\caption{Ganymede: Variations in orbital parameters over time introduce magnetic fluctuations at multiple frequencies beyond the Jupiter rotation and satellite orbital frequencies. }
\label{figure:frequenciesGanymede}
 \end{figure}

 \begin{figure}[t]
\centering
\includegraphics[width=30pc]{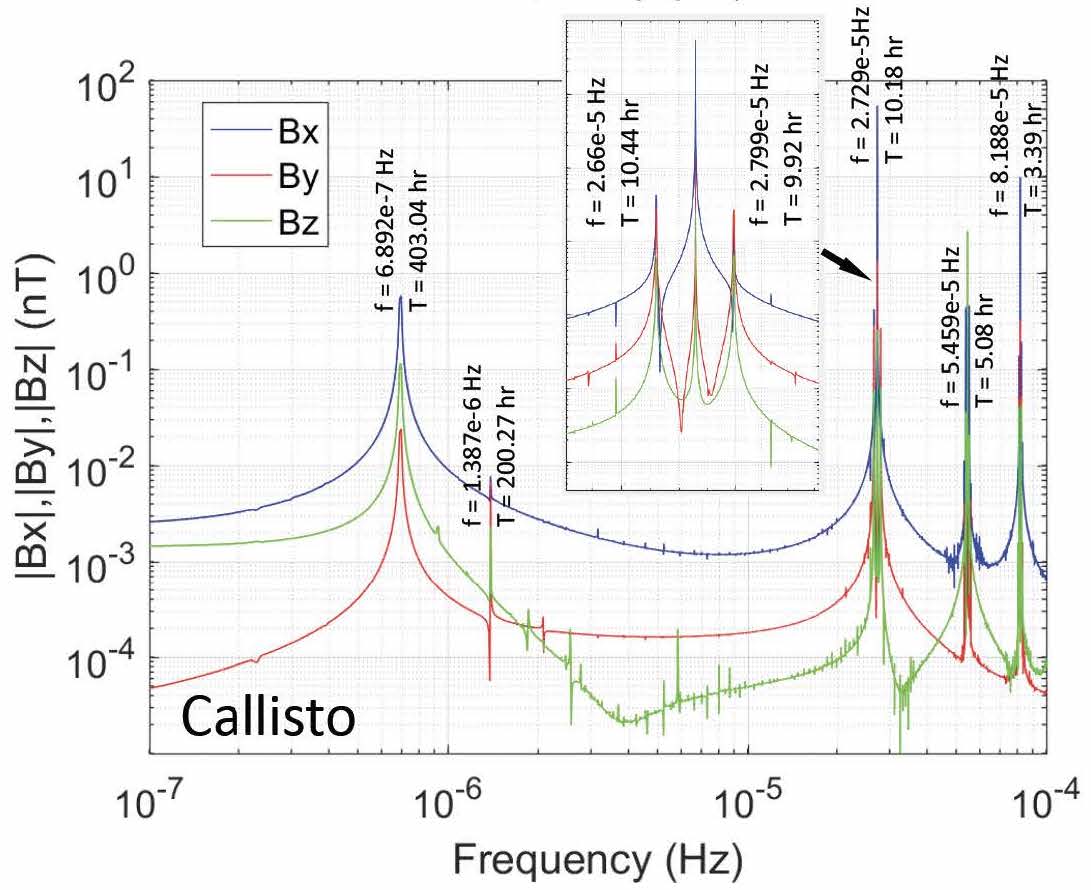}
\caption{Callisto: Variations in orbital parameters over time introduce magnetic fluctuations at multiple frequencies beyond the Jupiter rotation and satellite orbital frequencies.}
\label{figure:frequenciesCallisto}
 \end{figure}

\subsection{Electrical Conductivity in Adiabatic Galilean Oceans}\label{subsection:interiors}


Fluid temperature, pressure, and salt content determine the electrical conductivity of an aqueous solution, and thus dictate the magnetic induction responses of the Galilean oceans. The amplitude and phase of the magnetic fields induced by the oceans depend on the conductive properties of the oceans, which are influenced by the composition of the dissolved salts. With sufficient prior knowledge of the ice thickness and hints to the ocean's composition---for example, from geological and compositional measurements by the Europa Clipper \citep{buffington2017evolution}---magnetic induction studies can provide information on the amounts and compositions of the salts that link to global thermal and geochemical processes.  On Europa, the flux of surface-generated oxygen to the ocean may have created oxidizing (acidic) conditions \citep{hand2007empirical,pasek2012acidification,Vance_2016} permitting the presence of dissolved MgSO$_4$ in addition to NaCl \citep{zolotov2008oceanic,zolotov2009chemical}. 

\begin{figure}[h]
\centering
\includegraphics[width=30pc]{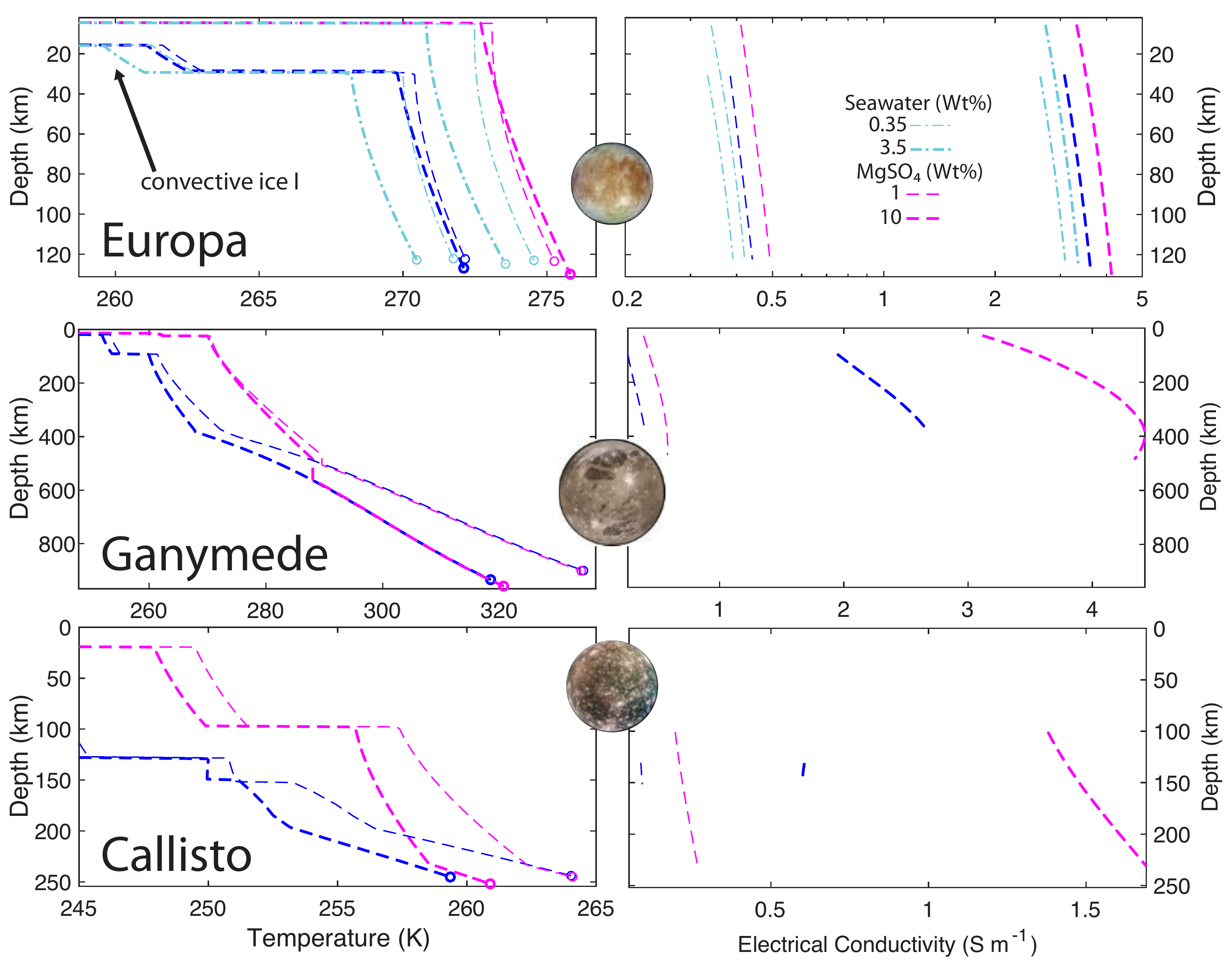}
\caption{Adiabatic ocean temperature (left) and electrical conductivity (right). Convecting oceans with MgSO$_4$ (dashed lines) are warmer. Standard seawater (mostly NaCl; dot--dashed lines) creates colder oceans and lower electrical conductivities. Thicker ice (blue), corresponds to colder adiabatic profiles in the underlying oceans, which also lowers electrical conductivity. Open and closed circles correspond to the inferred depth to the upper boundary of the silicate layer for the saline and pure water oceans, respectively. Conductivities in the liquid regions are several orders of magnitude larger than in the ice and rock. Adapted from \citet{Vance_2018}.}
\label{figure:conductivity}
 \end{figure}

Depth-dependent electrical conductivity can arise from melting or freezing at the ice--ocean interface, and from dissolution and precipitation within the ocean or at the water--rock interface. Even for  oceans with uniform salinity, as is typically assumed, conductivity will increase with depth along the ocean's convective adiabatic profile because the greater temperature and pressure increase the electrical conductivity. Figure~\ref{figure:conductivity} depicts this variation for Europa and Ganymede, based on forward models of \citet{Vance_2018} that use available thermodynamic and geophysical data to explore the influences of the ocean, rock layer, and any metallic core on the radial structures of known icy ocean worlds. For each ocean, we consider a nominal 10~wt\%~MgSO$_4$ salinity, as investigated in previous work. The published equation of state and electrical conductivity data are adequate for the pressures in the largest moon, Ganymede, up to 1.6~GPa \citep{Vance_2018}. The pressure conditions in Europa's ocean are low enough ($<200$~MPa) that the equation of state for seawater \citep{mcdougall2011getting} provides plausible values of conductivity for salinity of 35~ppt less. 
For Europa, the respective radial models of electrical conductivity for oceans containing seawater and MgSO$_4$  are consistent with compositions linked to chemically reducing and oxidizing model oceans cited above.

%
%

Radial conductivity profiles for Europa (Fig.~\ref{figure:conductivity}; top) illustrate the coupling to temperature and composition.  We consider ice thicknesses of 5 and 30~km (magenta and blue curves, respectively) as representative extremes. Seawater (dot--dashed lines), though less concentrated than the modeled composition of MgSO$_4$ (dashed lines), has a stronger melting point suppression, leading to an overall colder ocean for the same thickness of ice. Adiabats for pure water (solid lines) are shown for comparison. The lower temperature for seawater combines with the different electrical conductivity for the different dissolved ions to create distinct profiles unique to ocean composition and ice thickness (upper right). 

Larger Ganymede (Fig.~\ref{figure:conductivity}; bottom) also has distinct conductivity profiles for both ice thickness and ocean composition. They reveal an additional nuance to deep planetary oceans that can influence the induction response. Although electrical conductivity generally increases with depth, it begins to decrease at the greatest depths for the warm Ganymede ocean (right-most curve). This inflection occurs because the ocean achieves GPa+ pressures, at which the packing of water molecules begins to inhibit the charge exchange of the dissolved ions \citep{schmidt2017pressure}. 

Dense brines may also reside at the base of the high-pressure ices on Ganymede, and even between them \citep{Journaux_2013,Journaux_2017,Vance_2014,Vance_2018}. Although more detailed modeling of the coupled geochemical and geodynamic regimes is needed, this scenario seems consistent with recent simulations of two-phase convection in high-pressure ices \citep{Choblet_2017}. These simulations imply that fluids should occur at the water-rock interface through long periods of the evolution of even of large icy world containing high-pressure ices. If such a fluid layer exists under the high-pressure ice, it will create an induction response at low frequencies, as discussed below.

\subsection{Amplitude and Phase Lag of the Diffusive Response}\label{subsection:amplitudes}
\begin{figure}[h]
\centering
\includegraphics[width=30pc]{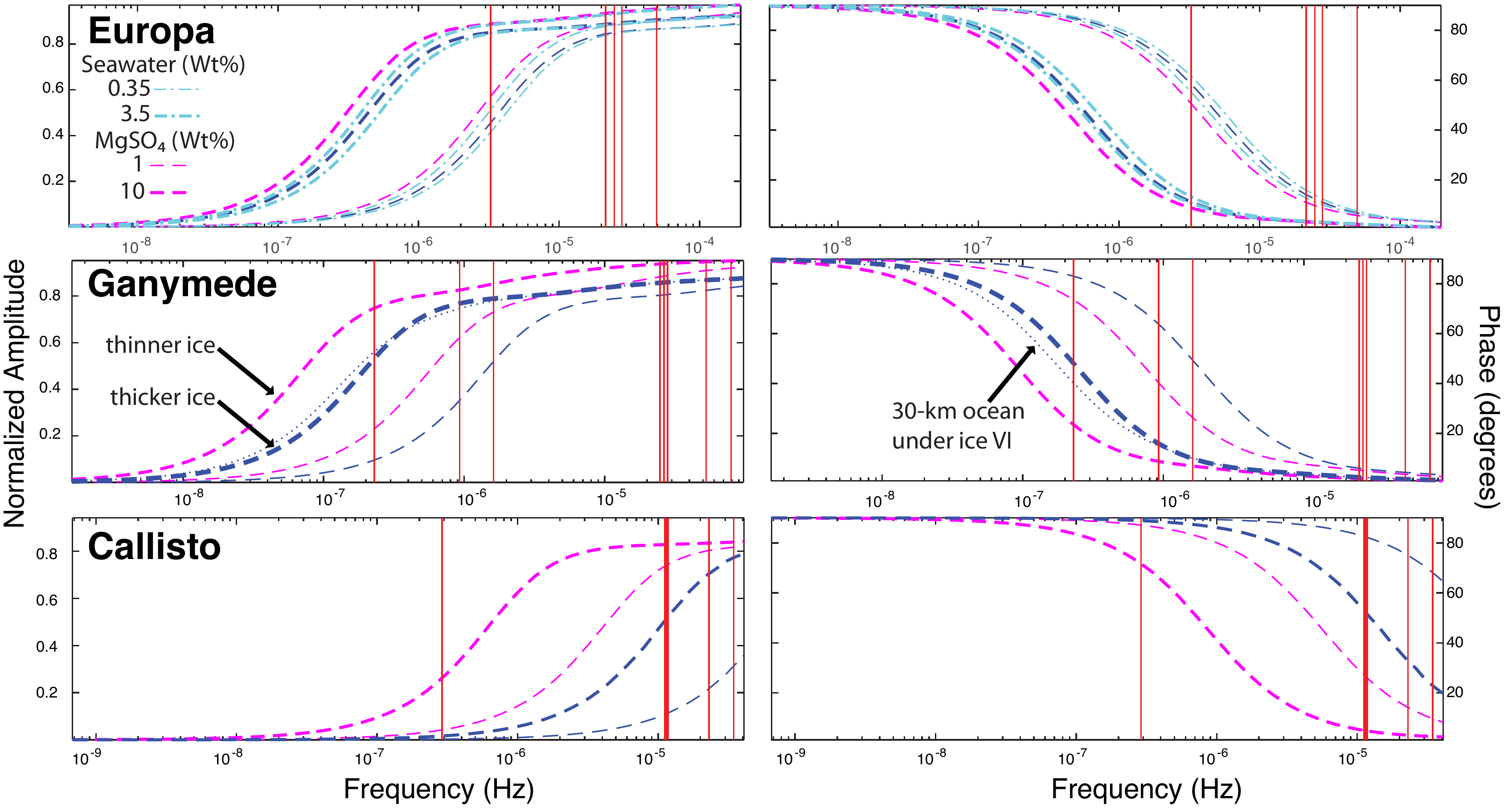}
\caption{Normalized magnetic induction amplitudes (left) and phases (right) for the conductivity profiles in Fig.~\ref{figure:conductivity}, at frequencies including the induction peaks noted in Fig.~\ref{figure:frequencies} (vertical red lines).}
\label{figure:amplitudes}
 \end{figure}

The normalized surface induction response for Europa, Ganymede, and Callisto, shown in Fig.~\ref{figure:amplitudes}, are based on the adiabatic ocean electrical conductivity profiles shown in Fig.~\ref{figure:conductivity}, assuming spherical symmetry (Section~\ref{section:inductionDerivation}). Warmer and thus thicker oceans (magenta curves) have larger amplitude responses, corresponding to overall higher values of the conductance. The induction signatures for the adiabatic ocean profile are nearly equal to those of oceans with uniform conductivity equal to the mean of the adiabatic model (Section~\ref{section:inductionDerivation}). These signatures differ, however, from those of an ocean with uniform conductivity based on the temperature and electrical conductivity at the ice--ocean interface. 

For Europa, the induction signatures for modeled oxidized (10~wt\% MgSO$_4$) and reduced (seawater) oceans are nearly identical in their amplitude responses. However, the two ocean models show phase separation of a few degrees at the orbital frequency of 3.6$\times10^{-6}$~Hz (85.23~hr period). 

Local enhancements in the ocean conductivity can have a discernible induction response. For Ganymede, we simulated a second ocean layer at the water--rock interface at a depth of 900~km, under 530~km of ice~VI \citep{Vance_2018}, modeled as a 10-km-thick high-conductivity region (20~S/m) corresponding to a nearly saturated MgSO$_4$ solution, consistent with \citep{hogenboom1995magnesium} and \citep{calvert1958determination}. 
The influence of such a layer (dotted lines in Fig.~\ref{figure:amplitudes}) is a $\sim$4\% increase in the amplitude response and a corresponding $\sim$7\% decrease in the phase response around 2.3$\times10^{-7}$~Hz. A $\sim$1\% decrease in amplitude is also seen at frequencies of 0.93$\times10^{-6}$~Hz and 1.6$\times10^{-6}$~Hz.

For Callisto, there is a small range of conditions under which oceans may be present. Salty oceans considered by \citet{Vance_2018} have thicknesses of 20 and 132~km. For the thinner ocean, a 96~km layer of high-pressure ice underlies the ocean. The depicted state is likely transient, as ice III is buoyant in the modeled 10wt\% MgSO$_4$ composition, and an upward snow effect should hasten the transfer of heat from the interior. Simulating a subsequent stage with ice~III above the ocean awaits improved thermodynamic data, and will be discussed in future work.  The present simulations illustrate the effect of the greater skin depth for the thicker and deeper ocean in terms of a higher amplitude response at lower frequencies and phase curve also shifted in the direction of lower frequencies.

\subsection{Mean Diffusive Response Relative to the Imposed Field}\label{subsection:residuals}

For the sake of comparing the passive induction responses of Europa, Ganymede and Callisto with fields induced by oceanic flows, we introduce the residual field, $B_R$. This quantity allows us to quickly examine the frequency dependent induction response for a given interior model, accounting for both the amplitude ($A$) and phase shift ($\phi$).  For the geometric mean frequency components of Jupiter's field ($|B|=\sqrt{B^2_x+B^2_y+B^2_z}$), we define $B_R$ as 
\begin{equation}
    B_R=|B|(\cos{\phi}-A)
\end{equation}
More information can be gained by examining the directional components of Jupiter's field (Figure \ref{figure:frequencies}). 

\begin{figure}[!tbp]
  \centering
  \includegraphics[width=1\textwidth]{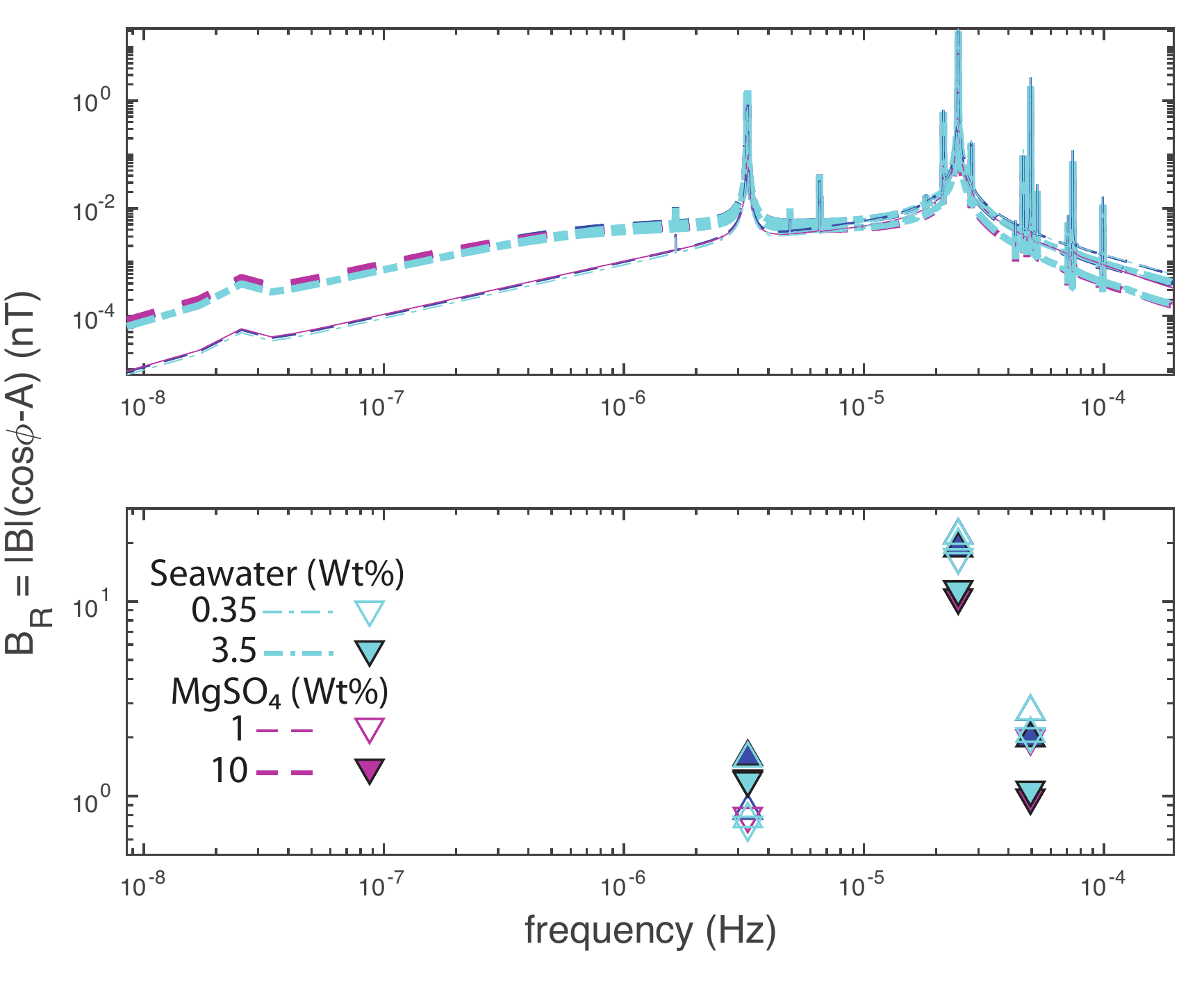}
  \caption{Europa:Residual field ($B_R$) of the diffusive induction response. Thick lines are higher salinities (10wt\% and 3.5wt\%, respectively) for oceans with aqueous MgSO$_4$ (magenta and blue $--$) and seawater (cyan dash-dot). Thinner lines are for oceans with 10\% of those concentrations. The lower pane shows responses at the strongest inducing frequencies in Figure~\ref{figure:frequencies}. Filled symbols are for the higher concentrations. Upward triangles are for thicker ice (30~km) and downward triangles are for thinner ice (5~km).}
\label{figure:EuropaResField}
\end{figure}

\begin{figure}[!tbp]
  \centering
  \includegraphics[width=1\textwidth]{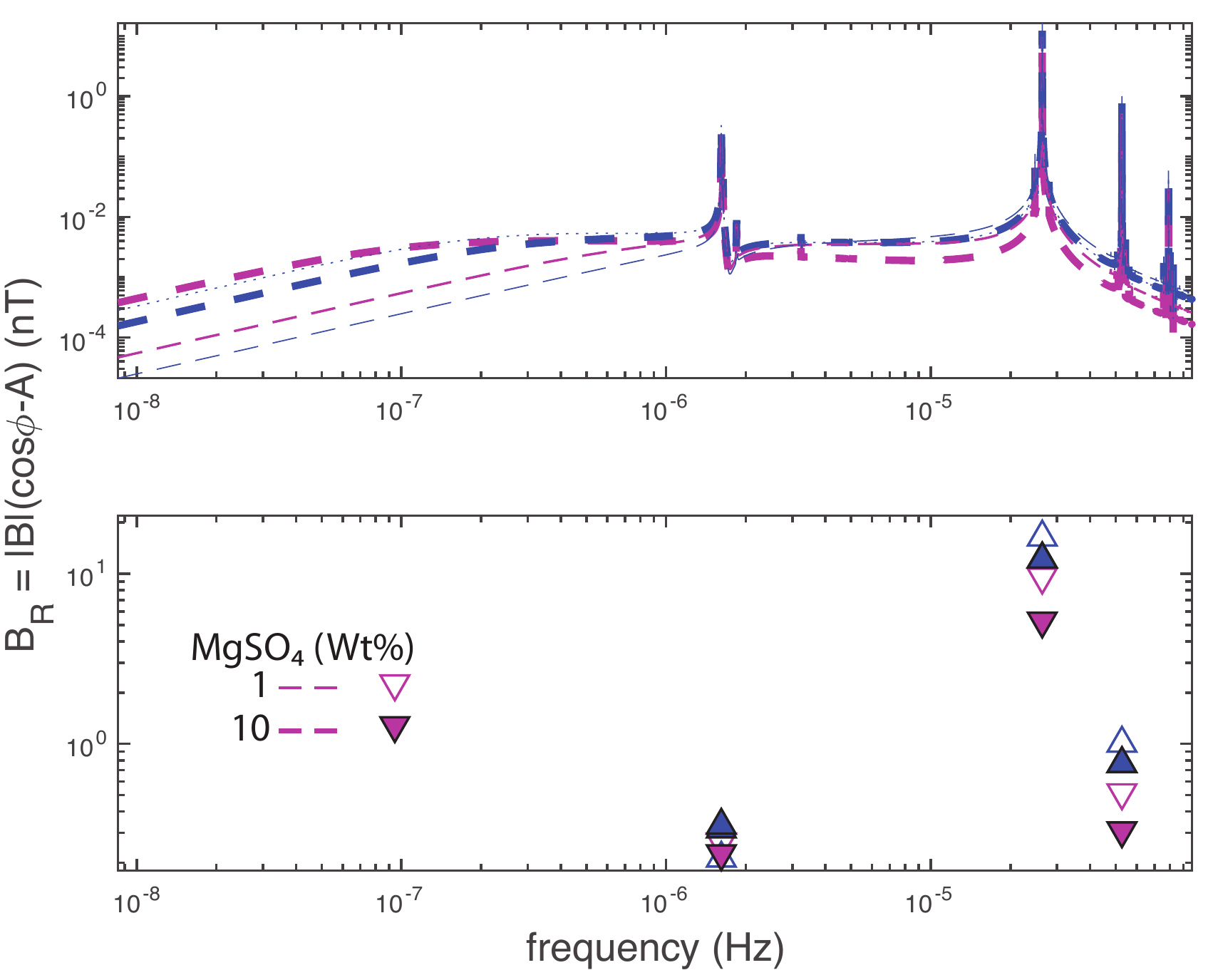}
  \caption{Ganymede: Residual field ($B_R$) of the diffusive induction response. Thick lines are higher salinities (10wt\%) for oceans with aqueous MgSO$_4$ (magenta and blue $--$). Thinner lines are for oceans with 1wt\% MgSO$_4$. The dotted line is for the case with a 30-km-thick  oceanic layer underneath the high-pressure ice. The lower pane shows responses at the strongest inducing frequencies in Figure~\ref{figure:frequencies}. Filled symbols are for the higher concentrations. Upward triangles are for thicker ice ($\sim100$~km) and downward triangles are for thinner ice ($\sim30$~km)}
\label{figure:GanymedeResField}
\end{figure}

\begin{figure}[!tbp]
  \centering
  \includegraphics[width=1\textwidth]{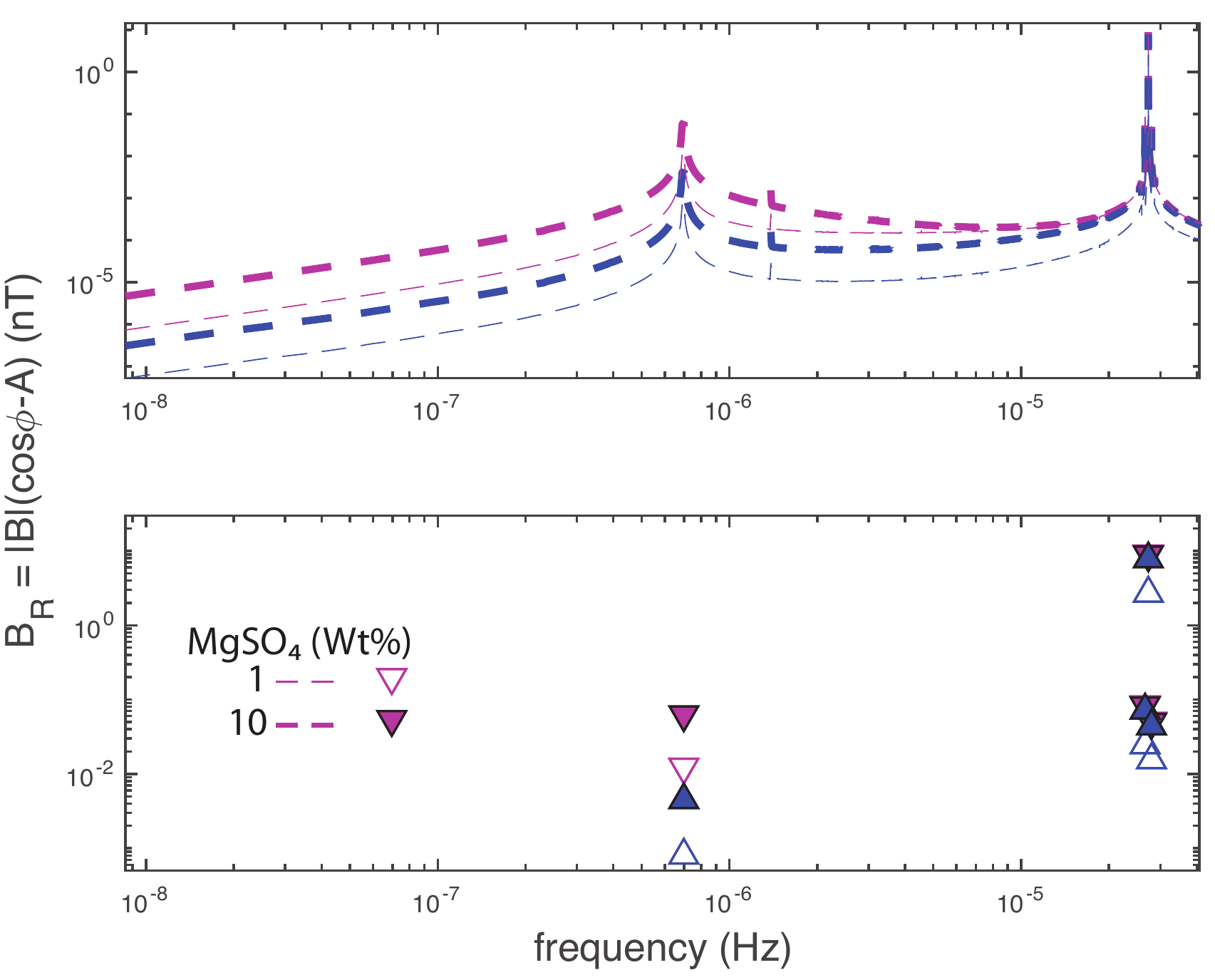}
  \caption{Callisto: Residual field ($B_R$) of the diffusive induction response. Thick lines are higher salinities (10wt\%) for oceans with aqueous MgSO$_4$ (magenta and blue $--$). Thinner lines are for oceans with 1wt\% MgSO$_4$. The lower pane shows responses at the strongest inducing frequencies in Figure~\ref{figure:frequencies}. Filled symbols are for the higher concentrations. Upward triangles are for thicker ice ($\sim130$~km) and downward triangles are for thinner ice ($\sim100$~km).}
\label{figure:CallistoResField}
\end{figure}

Figures~\ref{figure:EuropaResField}, \ref{figure:GanymedeResField}, and \ref{figure:CallistoResField} show the spectra of residual fields for Europa, Ganymede, and Callisto, respectively. Subpanels in each figure isolate the peak responses at the main driving frequencies shown in Figure \ref{figure:frequencies}. Tables \ref{table:inducedEuropa}, \ref{table:inducedGanymede}, and \ref{table:inducedCallisto} include the corresponding data. Figures~S1-S3 illustrate possible errors arising from analyses assuming a uniform conductivity of the ocean. They plot the deviations (in percent) between the residual fields ($B_R$) of the adiabatic oceans (Figure~\ref{figure:conductivity}) and the equivalent responses obtained by giving the oceans uniform conductivity, either as the equivalent mean value or the value at the top of the ocean (i.e. at the ice--ocean interface). 

\begin{table}[h]
\begin{center}
\begin{tabular}{|c|c|c|c|c|c c c |}
\hline
 & $T_{b}$ & $T_{mean}$ &  $D_{I}$  & $D_{ocean}$        &  \multicolumn{3}{c|}{$B_R$} \\
 & (K)     & (K)       &    (km)     & (km)               & \multicolumn{3}{c|}{(nT)} \\
\hline
\hline
Europa &         &          &                         \multicolumn{2}{c}{$f$ ($\times 10^{-6}$Hz)}   & \textbf{3.25} & \textbf{24.73} & \textbf{49.46} \\
\hline
\textbf{MgSO$_{4}$ 1Wt\%} &270.4 &271.5 &31 &120 &0.841 &21.862 &2.715 \\
$<\sigma>=$ 0.4227 S m$^{-1}$ &270.4 &271.5 &31 &120 &0.823 &21.417 &2.654 \\
$\sigma_{top}=$ 0.3847 S m$^{-1}$ &270.4 &271.5 &31 &120 &0.769 &21.304 &2.650 \\
\hline
 &273.1 &274.3 &6 &147 &0.791 &16.892 &1.980 \\
$<\sigma>=$ 0.4640 S m$^{-1}$ &273.1 &274.3 &6 &147 &0.755 &15.964 &1.900 \\
$\sigma_{top}=$ 0.4107 S m$^{-1}$ &273.1 &274.3 &6 &147 &0.702 &16.122 &1.928 \\
\hline
\textbf{MgSO$_{4}$ 10Wt\%} &269.8 &271.3 &30 &127 &1.591 &18.741 &1.983 \\
$<\sigma>=$ 3.4478 S m$^{-1}$ &269.8 &271.3 &30 &127 &1.539 &18.234 &1.961 \\
$\sigma_{top}=$ 3.0763 S m$^{-1}$ &269.8 &271.3 &30 &127 &1.536 &18.686 &2.008 \\
\hline
 &272.7 &274.5 &5 &154 &1.233 &10.477 &0.982 \\
$<\sigma>=$ 3.8547 S m$^{-1}$ &272.7 &274.5 &5 &154 &1.167 &9.800 &0.935 \\
$\sigma_{top}=$ 3.3197 S m$^{-1}$ &272.7 &274.5 &5 &154 &1.173 &10.634 &1.000 \\
\hline
\textbf{Seawater 0.35165 Wt\%} &270.0 &271.1 &31 &120 &0.763 &21.749 &2.719 \\
$<\sigma>=$ 0.3734 S m$^{-1}$ &270.0 &271.1 &31 &120 &0.746 &21.112 &2.645 \\
$\sigma_{top}=$ 0.3339 S m$^{-1}$ &270.0 &271.1 &31 &120 &0.684 &21.026 &2.636 \\
\hline
 &272.5 &273.6 &6 &146 &0.712 &16.850 &2.029 \\
$<\sigma>=$ 0.3945 S m$^{-1}$ &272.5 &273.6 &6 &146 &0.678 &16.046 &1.926 \\
$\sigma_{top}=$ 0.3415 S m$^{-1}$ &272.5 &273.6 &6 &146 &0.614 &15.921 &1.947 \\
\hline
\textbf{Seawater 3.5165 Wt\%} &268.2 &269.7 &31 &122 &1.559 &19.524 &2.091 \\
$<\sigma>=$ 2.9548 S m$^{-1}$ &268.2 &269.7 &31 &122 &1.523 &18.989 &2.052 \\
$\sigma_{top}=$ 2.6476 S m$^{-1}$ &268.2 &269.7 &31 &122 &1.510 &19.349 &2.098 \\
\hline
 &270.8 &272.3 &5 &148 &1.205 &11.538 &1.079 \\
$<\sigma>=$ 3.1457 S m$^{-1}$ &270.8 &272.3 &5 &148 &1.138 &10.805 &1.024 \\
$\sigma_{top}=$ 2.7346 S m$^{-1}$ &270.8 &272.3 &5 &148 &1.140 &11.350 &1.068 \\
\hline
\end{tabular}
\end{center}
\caption{Europa: Residual fields ($B_R$) at the main inducing frequencies in Fig~\ref{figure:frequencies}. For the different ocean compositions and thicknesses of the upper ice~I lithosphere ($D_I$; Figure~\ref{figure:conductivity}, the adiabatic response is given first, followed by the response for the ocean with uniform conductivity set to the mean of the adiabatic ocean ($\langle\sigma\rangle$), and then for the case with uniform conductivity set to the value at the ice-ocean interface ($\sigma_{top}$).} 
\label{table:inducedEuropa}
\end{table}%

\begin{table}[h]
\begin{center}
\begin{tabular}{|c|c|c|c|c|c c c |}
\hline
 & $T_{b}$ & $T_{mean}$ &  $D_{I}$  & $D_{ocean}$        &  \multicolumn{3}{c|}{$B_R$} \\
 & (K)     & (K)       &    (km)     & (km)               & \multicolumn{3}{c|}{(nT)} \\
\hline
\hline
Ganymede &         &          &                         \multicolumn{2}{c}{$f$ ($\times 10^{-6}$Hz)}   & \textbf{1.62} & \textbf{26.37} & \textbf{52.74} \\
\hline
\textbf{MgSO$_4$ 1Wt\%} &270.7 &279.0 &25 &442 &0.265 &9.580 &0.517 \\
$<\sigma>=$ 0.5166 S m$^{-1}$ &270.7 &279.0 &25 &442 &0.243 &8.753 &0.477 \\
$\sigma_{top}=$ 0.3890 S m$^{-1}$ &270.7 &279.0 &25 &442 &0.229 &9.601 &0.516 \\
\hline
 &261.5 &266.1 &93 &272 &0.212 &16.389 &1.007 \\
$<\sigma>=$ 0.3295 S m$^{-1}$ &261.5 &266.1 &93 &272 &0.203 &15.626 &0.967 \\
$\sigma_{top}=$ 0.2608 S m$^{-1}$ &261.5 &266.1 &93 &272 &0.175 &15.906 &0.999 \\
\hline
\textbf{MgSO$_4$ 10Wt\%} &270.1 &278.2 &28 &455 &0.226 &5.286 &0.309 \\
$<\sigma>=$ 4.0541 S m$^{-1}$ &270.1 &278.2 &28 &455 &0.209 &4.991 &0.290 \\
$\sigma_{top}=$ 3.1056 S m$^{-1}$ &270.1 &278.2 &28 &455 &0.226 &5.325 &0.306 \\
\hline
 &260.0 &263.5 &96 &282 &0.316 &12.202 &0.762 \\
$<\sigma>=$ 2.3476 S m$^{-1}$ &260.0 &263.5 &96 &282 &0.304 &11.919 &0.750 \\
$\sigma_{top}=$ 1.9483 S m$^{-1}$ &260.0 &263.5 &96 &282 &0.304 &12.174 &0.761 \\
\hline
30~km 20~S~m$^{-1}$  layer  &260.0 &263.5 &96 &282 &0.332 &12.156 &0.765 \\
\hline
\end{tabular}
\end{center}
\caption{Ganymede:  Residual fields ($B_R$) at the main inducing frequencies in Fig~\ref{figure:frequencies}. For the different ocean compositions and thicknesses of the upper ice~I lithosphere ($D_I$; Figure~\ref{figure:conductivity}, the adiabatic response is given first, followed by the response for the ocean with uniform conductivity set to the mean of the adiabatic ocean ($\langle\sigma\rangle$), and then for the case with uniform conductivity set to the value at the ice-ocean interface ($\sigma_{top}$).}
\label{table:inducedGanymede}
\end{table}%

\begin{table}[h]
\begin{center}
\begin{tabular}{|c|c|c|c|c|c c c c |}
\hline
 & $T_{b}$ & $T_{mean}$ &  $D_{I}$  & $D_{ocean}$        &  \multicolumn{4}{c|}{$B_R$} \\
 & (K)     & (K)       &    (km)     & (km)               & \multicolumn{4}{c|}{(nT)} \\
\hline
\hline
Callisto &         &          &                         \multicolumn{2}{c}{$f$ ($\times 10^{-6}$Hz)}   & \textbf{0.69} & \textbf{26.60} & \textbf{27.29} & \textbf{27.99} \\
\hline
\textbf{MgSO$_{4}$ 1Wt\%} &257.4 &259.6 &99 &132 &0.012 &0.085 &9.201 &0.052 \\
$<\sigma>=$ 0.2307 S m$^{-1}$ &257.4 &259.6 &99 &132 &0.012 &0.084 &8.990 &0.050 \\
$\sigma_{top}=$ 0.1965 S m$^{-1}$ &257.4 &259.6 &99 &132 &0.010 &0.083 &8.926 &0.050 \\
\hline
 &250.8 &250.9 &128 &21 &0.001 &0.024 &2.688 &0.015 \\
$<\sigma>=$ 0.0895 S m$^{-1}$ &250.8 &250.9 &128 &21 &0.001 &0.025 &2.740 &0.016 \\
$\sigma_{top}=$ 0.0874 S m$^{-1}$ &250.8 &250.9 &128 &21 &0.001 &0.024 &2.689 &0.015 \\
\hline
\textbf{MgSO$_{4}$ 10Wt\%} &255.7 &256.9 &99 &130 &0.063 &0.083 &8.875 &0.050 \\
$<\sigma>=$ 1.5256 S m$^{-1}$ &255.7 &256.9 &99 &130 &0.062 &0.082 &8.763 &0.049 \\
$\sigma_{top}=$ 1.3789 S m$^{-1}$ &255.7 &256.9 &99 &130 &0.058 &0.082 &8.822 &0.049 \\
\hline
 &250.0 &251.5 &129 &18 &0.004 &0.072 &7.778 &0.044 \\
$<\sigma>=$ 0.6025 S m$^{-1}$ &250.0 &251.5 &129 &18 &0.005 &0.072 &7.781 &0.044 \\
$\sigma_{top}=$ 0.6062 S m$^{-1}$ &250.0 &251.5 &129 &18 &0.005 &0.072 &7.790 &0.044 \\
\hline
\end{tabular}
\end{center}
\caption{Callisto: Residual fields ($B_R$) at the main inducing frequencies in Fig~\ref{figure:frequencies}. For the different ocean compositions and thicknesses of the upper ice~I lithosphere ($D_I$; Figure~\ref{figure:conductivity}, the adiabatic response is given first, followed by the response for the ocean with uniform conductivity set to the mean of the adiabatic ocean ($\langle\sigma\rangle$), and then for the case with uniform conductivity set to the value at the ice-ocean interface ($\sigma_{top}$).}
\label{table:inducedCallisto}
\end{table}%

\section{Magnetic Induction from Oceanic Fluid Flows}\label{section:flows}

Another component of the induced magnetic response might occur in the icy Galilean satellites, arising not from Jupiter's changing magnetic field, but from charges moving with oceanic fluid flows. Such induced magnetic fields are typically neglected because they are expected to be relatively weak. On Earth, ocean currents induce fields on the order of 100~nT in a background field of about 40,000~nT; these fields are observable by space-based magnetometers and have been used to monitor ocean currents \citep{ConstableConstable04,TylerEA03}. If there are oceanic flow-driven induction signals present in the icy Galilean satellites, and if the spatial or temporal structures of these induction signals allow them to be separated from the contributions driven by variations in Jupiter's magnetic field, it would permit characterization of the ocean flows themselves as has been done for the Earth's ocean 
\citep[e.g.,][]{Chave83,TylerEA03,GrayverEA16,Minami17}. Conversely, if such induced signals are present but the analysis does not accommodate that fact, then the recovered electrical conductivity estimates will be biased and inaccurate. 

While \citet{Tyler11} discusses the possibility of magnetic remote sensing to detect resonant ocean tides on Europa in the limits of shallow water equations and thin-shell electrodynamics, we are not aware of any studies that have examined magnetic induction signatures due to other flows or for other satellites \citep[e.g.,][]{LemasquerierEA17,GissingerPetitdemange19,RoviraNavarroEA19,Soderlund19}. 
Here, we focus on global fluid motions that may be driven by convection within the oceans of Europa, Ganymede, and Callisto, followed by estimates of the induction response that may be expected from these flows.


\subsection{Oceanic Fluid Motions}\label{subsection:motions}

\begin{figure}[h]
\centering
\includegraphics[width=33pc]{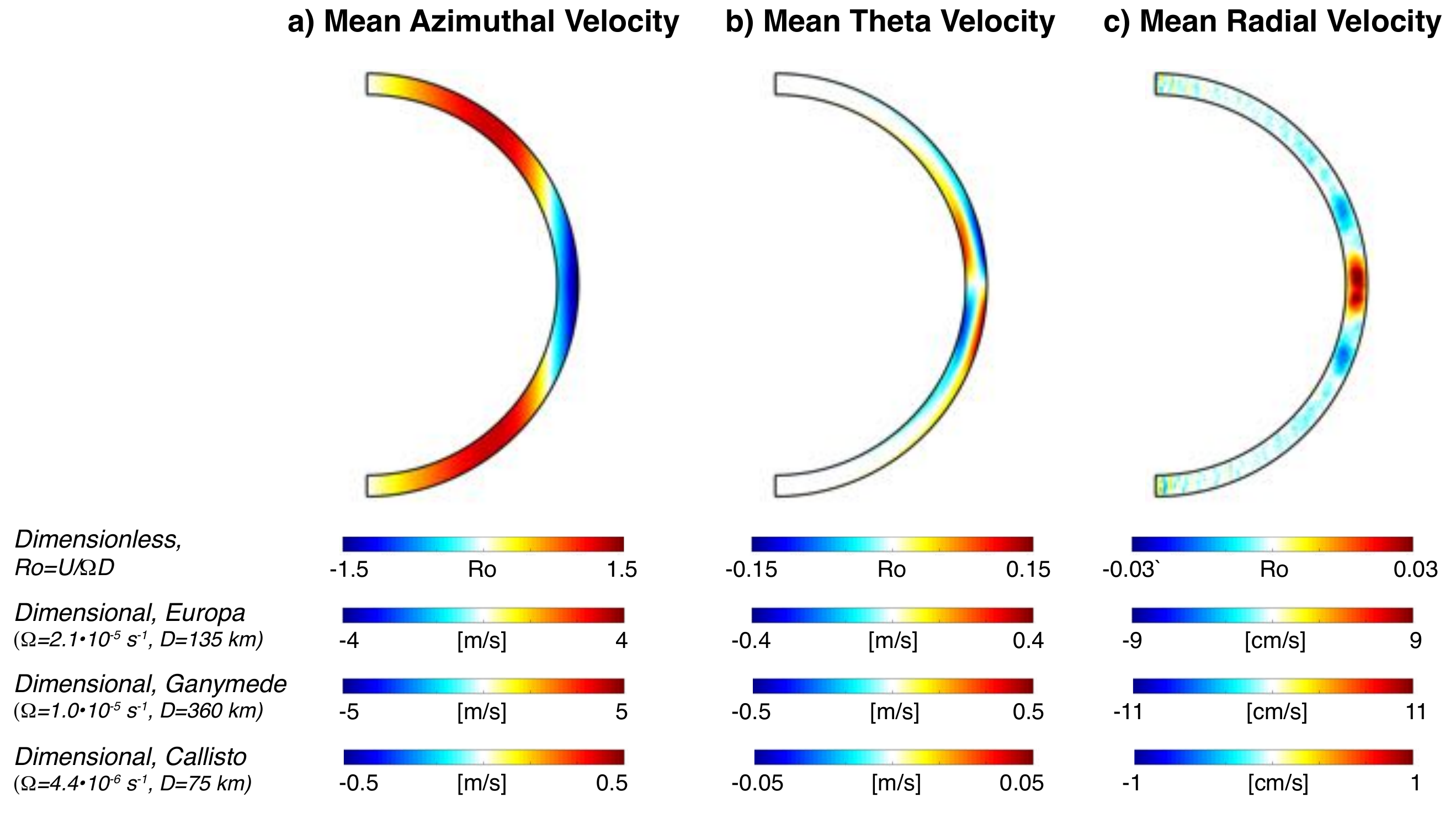}
\caption{Mean flow fields in our nominal global ocean model from \citet{Soderlund19}, averaged over 18 planetary rotations and all longitudes. {\bf a)} Zonal velocity field where red denotes prograde flows and blue denotes retrograde flows. {\bf b)} Theta velocity field where red denotes away from the north pole and blue denotes toward the north pole. {\bf c)} Radial velocity field where red denotes upwelling flows and blue denotes downwelling flows.}%
\label{fig:oceanflows}
 \end{figure}

The majority of ocean circulation studies have focused on hydrothermal plumes at Europa, with global models being developed relatively recently \citep{VanceGoodman09,soderlund2014ocean,Soderlund19}. Thermal convection in Europa's ocean is expected in order to efficiently transport heat from the deeper interior that arises primarily from radiogenic and tidal heating in the mantle. Moreover, by estimating the extent to which rotation will organize the convective flows, Europa's ocean was predicted to have quasi-three-dimensional turbulence \citep{soderlund2014ocean,Soderlund19}. As shown in Figure~\ref{fig:oceanflows}, this turbulence generates three-jet zonal flows with retrograde (westward) flow at low latitudes, prograde (eastward) flow at high latitudes, and meridional overturning circulation. Upwelling at the equator and downwelling at middle to high latitudes from this circulation effectively forms a Hadley-like cell in each hemisphere.

Application of these calculations to Ganymede suggests convection is expected within its ocean as well and may have similar convective flows, although there is significantly more uncertainty in the predicted convective regime \citep{Soderlund19}. Convection in Callisto's potential ocean may be in the double-diffusive regime if the ocean's composition is nearly saturated \citep{Vance_2018}. However, considering thermal convection as an upper bound, application of the scaling arguments in \citet{Soderlund19} to Callisto suggest similar ocean flows may be expected here as well.

The nominal ocean model shown in Figure~\ref{fig:oceanflows} is, therefore, applicable to all three ocean worlds considered here. As described in \citet{Soderlund19}, the model was carried out using the MagIC code \citep{Wicht02} with the SHTns library for the spherical harmonics transforms \citep{Schaeffer13} and is characterized by the following dimensionless input parameters: shell geometry $\chi=r_i/r_o=0.9$, Prandtl number $Pr=\nu / \kappa=1$, Ekman number $E=\nu / \Omega D^2=3.0 \times 10^{-4}$, and Rayleigh number $Ra=\alpha g \Delta T D^3 / \nu \kappa$, where $r_i$ and $r_o$ are the inner and outer radii of the ocean, $D=r_o-r_i$ is ocean thickness, $\Omega$ is rotation rate, $\nu$ is kinematic viscosity, $\kappa$ is thermal diffusivity, $\alpha$ is thermal expansivity, $g$ is gravitational acceleration, and $\Delta T$ is superadiabatic temperature contrast. The boundaries are impenetrable, stress-free, and isothermal. 

The model outputs, such as the velocity field, are also non-dimensional. For example, the Rossby number $Ro=U/\Omega D$ is the ratio of rotational $\Omega^{-1}$ to inertial $D/U$ timescales that allows the dimensional flow speeds to be determined: $U = \Omega D Ro$ using ocean thickness $D$ as the length scale and rotation rates $\Omega=[2.1 \times 10^{-5}, 1.0 \times 10^{-5}, 4.4 \times 10^{-6}]$ s$^{-1}$ for Europa, Ganymede, and Callisto, respectively. Following Table~\ref{table:inducedEuropa}, Europan ocean thicknesses of $120-154$~km are considered. This range of liquid ocean thicknesses extends to $272-455$~km for Ganymede (Table~\ref{table:inducedGanymede}) and $18-132$~km for Callisto (Table~\ref{table:inducedCallisto}), given the larger uncertainties on their internal structures. We therefore assume the following mean parameter values in Figure~\ref{fig:oceanflows}: $D_{Europa}=135$ km, $D_{Ganymede} = 360$ km, and $D_{Callisto} = 75$ km, with the ranges considered in Table~\ref{table:flow}. Flows are fastest for Ganymede and Europa, where the zonal jets can reach m/s speeds, the mean latitudinal flows have peak speeds of tens of cm/s, and the mean radial flows are $\sim10$~cm/s. 

\subsection{Generation of Induced Magnetic Fields} \label{subsection:flowinduction}

The magnetic induction equation can be used to estimate the components of the magnetic field ${\bf B}$ induced by ocean currents with velocity ${\bf u}$ and those arising from changes in the externally imposed field:
\begin{equation}
    \frac{\partial {\bf B}}{\partial t} = \nabla \times ({\bf u} \times {\bf B}) - \nabla \times (\eta \nabla \times {\bf B})
    \label{eqn:induction_full}
\end{equation}
where $\eta=(\mu_0 \sigma)^{-1}$ is the magnetic diffusivity. Here, the first term represents the evolution of the magnetic field, the second term represents magnetic induction, and the third term represents magnetic diffusion.

 Neglecting variations in oceanic electrical conductivity with depth and assuming an incompressible fluid, equation~\ref{eqn:induction_full} simplifies to
\begin{equation}
    \frac{\partial {\bf B}}{\partial t} = ({\bf B} \cdot \nabla ) {\bf u} - ({\bf u} \cdot \nabla) {\bf B} + \eta \nabla^2 {\bf B},
    \label{eqn:induction_short}
\end{equation}
after also expanding the induction term and utilizing $\nabla \cdot {\bf B} = 0$ and $\nabla \cdot {\bf u} = 0$. Let us decompose the total magnetic field into jovian imposed ${\bf F}$ and the satellite's induced ${\bf b}$ field components:
\begin{equation}
        {\bf B} = {\bf F} + {\bf b}
\end{equation}
with $|{\bf F}| \gg |{\bf b}|$. The induction equation then becomes 
\begin{equation}
    \frac{\partial {\bf b}}{\partial t} = - \frac{\partial {\bf F}}{\partial t} + ({\bf F} \cdot \nabla ) {\bf u} - ({\bf u} \cdot \nabla) ({\bf F} +{\bf b}) + \eta \nabla^2 ({\bf F} + {\bf b})
    \label{eqn:induction_decomp}
\end{equation}
Here, the first term is the evolution of the induced magnetic field, the second term is induction due to variations in Jupiter's magnetic field, the third term is induction due to oceanic fluid motions, the fourth term is advection of the field by ocean flows, and the fifth and sixth terms are diffusion of the Jovian and induced fields.

Let us next assume that the Jovian field can be approximated by ${\bf F}=F_o {\bf \hat{z}}$, where $F_o$ is constant and homogeneous and ${\bf \hat{z}}$ is aligned with the rotation axis, in which case equation~\ref{eqn:induction_decomp} further simplifies to:
\begin{equation}
    \frac{\partial {\bf b}}{\partial t} =  F_o \frac{\partial {\bf u}}{\partial z} - ({\bf u} \cdot \nabla) {\bf b} + \eta \nabla^2 {\bf b}.
    \label{eqn:induction_decomp_short}
\end{equation}
We will also focus on the quasi-steady induction signal generated by ocean flows rather than the rapidly varying contribution that could be difficult to distinguish from other magnetic field perturbations. Towards this end, the induced magnetic field and velocity fields are decomposed into mean and fluctuating components: ${\bf b} = \overline{\bf b} + {\bf b'}$ and ${\bf u} = \overline{\bf u} + {\bf u'}$. Inserting this into equation~\ref{eqn:induction_decomp_short} and using Reynolds averaging yields
\begin{equation}
    \frac{\partial \overline{\bf b}}{\partial t} =  F_o \frac{\partial \overline{\bf u}}{\partial z} - (\overline{\bf u} \cdot \nabla) \overline{\bf b} - \overline{({\bf u'} \cdot \nabla) {\bf b'}} + \eta \nabla^2 \overline{\bf b}.
    \label{eqn:induction_decomp_Reynolds}
\end{equation}
Next, we focus on the radial and latitudinal components because the zonal flow ($\overline{u_{\phi}}$) is nearly invariant in the z-direction (Figure~\ref{fig:oceanflows}a), noting also that azimuthally oriented (toroidal) magnetic fields would not be detectable by spacecraft:
\begin{equation}
    \frac{\partial \overline{b_r}}{\partial t} =  F_o \frac{\partial \overline{u_r}}{\partial z} - (\overline{{\bf u}} \cdot \nabla) \overline{b_r} - \overline{({\bf u'} \cdot \nabla) b_r'} + \eta \nabla^2 \overline{b_r}
    \label{eqn:induction_decomp_r}
\end{equation}
\begin{equation}
    \frac{\partial \overline{b_{\theta}}}{\partial t} =  F_o \frac{\partial \overline{u_{\theta}}}{\partial z} - (\overline{{\bf u}} \cdot \nabla) \overline{b_{\theta}} - \overline{({\bf u'} \cdot \nabla) b_{\theta}'} + \eta \nabla^2 \overline{b_{\theta}}
    \label{eqn:induction_decomp_theta}
\end{equation}
Using simple scaling arguments, the second and third terms on the right sides are likely small compared to the first term since $|F| \gg |b|$ (assuming similar characteristic flow speeds and length scales) such that 
\begin{equation}
    \frac{\partial \overline{b_r}}{\partial t} \approx  F_o \frac{\partial \overline{u_r}}{\partial z} + \eta \nabla^2 \overline{b_r}
    \label{eqn:induction_decomp_rshort}
\end{equation}
\begin{equation}
    \frac{\partial \overline{b_{\theta}}}{\partial t} \approx  F_o \frac{\partial \overline{u_{\theta}}}{\partial z} + \eta \nabla^2 \overline{b_{\theta}}.
    \label{eqn:induction_decomp_thetashort}
\end{equation}

In the steady state limit and approximating the gradient length scales as $D$ and flow speeds as $U_r$ and $U_{\theta}$, the magnetic fields induced by ocean currents can be estimated as:
\begin{equation}
\frac{F_o U_r}{D} \sim \frac{\eta b_r}{D^2} \mbox{ such that } b_r \sim \frac{F_o U_r D}{\eta} = \mu_o \sigma D U_r F_o
\end{equation}
\begin{equation}
\frac{F_o U_{\theta}}{D} \sim \frac{\eta b_{\theta}}{D^2} \mbox{ such that } b_{\theta} \sim \frac{F_o U_{\theta} D}{\eta} = \mu_o \sigma D U_{\theta} F_o.
\end{equation}
The resulting induced magnetic fields are then stronger for larger electrical conductivities, ocean thicknesses, flow velocities, and satellites closer to the host planet, since $F_o$ decreases with distance.

Table~\ref{table:flow} summarizes the ambient Jovian conditions at Europa, Ganymede, and Callisto as well as the relevant characteristics of their oceans, and the computed upper bounds on the induced magnetic field strengths. Here, we assume flow speeds typical of the global, steady overturning cells due to their temporal persistence and large spatial scale, which we hypothesize will produce the strongest induced magnetic signature that would be detectable at spacecraft altitudes. We find that the theta magnetic field components are larger than the radial components by roughly a factor of five, reaching $\sim 200$ nT for both Europa and Ganymede (higher salt content, thinner ice shell models); estimates can be an order of magnitude weaker in the lower salt content, thicker ice shell models). The radial variations correspond to signals up to 33\% (Ganymede) and 8\% (Europa) of the ambient Jovian field, which could be detectable with future missions. The signature at Callisto is small ($\lesssim 1$ nT). In addition, we predict the fields to be strongest near the equator where large vertical gradients in the convective flows exist (Figure~\ref{fig:oceanflows}b-c).

\begin{table}[h]
\centering 
\begin{tabular}{l|ccccc|cc}
\hline 
         & $\sigma$  & $D$   & $U_r$ & $U_{\theta}$ & $F_o$ & $b_r$ & $b_{\theta}$ \\
         & [S/m]     & [km]  & [m/s] & [m/s]        & [nT]  & [nT]  & [nT]         \\
\hline
Europa & & & & & & & \\
\hline
MgSO$_4$ 1 Wt\%,    Thicker ice shell  & 0.4   & 120   & 0.08  & 0.38 & 420  & 2   & 10   \\
MgSO$_4$ 1 Wt\%,    Thinner ice shell  & 0.5   & 147   & 0.09  & 0.46 & 420  & 3   & 18   \\
MgSO$_4$ 10 Wt\%,   Thicker ice shell  & 3.4   & 127   & 0.08  & 0.40 & 420  & 18  & 91   \\
MgSO$_4$ 10 Wt\%,   Thinner ice shell  & 3.9   & 154   & 0.10  & 0.49 & 420  & 32  & 155  \\
Seawater 0.35 Wt\%, Thicker ice shell  & 0.4   & 120   & 0.08  & 0.38 & 420  & 2   & 10   \\
Seawater 0.35 Wt\%, Thinner ice shell  & 0.4   & 146   & 0.09  & 0.46 & 420  & 3   & 14   \\
Seawater 3.5 Wt\%,  Thicker ice shell  & 3.0   & 122   & 0.08  & 0.38 & 420  & 15  & 73   \\
Seawater 3.5 Wt\%,  Thinner ice shell  & 3.1   & 148   & 0.09  & 0.47 & 420  & 22  & 114  \\
\hline
Ganymede & & & & & & & \\
\hline
MgSO$_4$ 1 Wt\%, Thicker ice shell  & 0.3   & 272   & 0.08  & 0.41 & 120  & 1    & 5    \\ 
MgSO$_4$ 1 Wt\%, Thinner ice shell  & 0.5   & 442   & 0.13  & 0.66 & 120  & 4    & 22    \\ 
MgSO$_4$ 10 Wt\%, Thicker ice shell & 2.3   & 282   & 0.08  & 0.42 & 120  & 8    & 41    \\ 
MgSO$_4$ 10 Wt\%, Thinner ice shell & 4.1   & 455   & 0.14  & 0.68 & 120  & 39   & 191    \\ 
\hline
Callisto & & & & & & & \\
\hline
MgSO$_4$ 1 Wt\%,  Thicker ice shell & 0.09 & 21    & 0.003 & 0.01 & 35   & $\ll 1$ & $\ll 1$ \\ 
MgSO$_4$ 1 Wt\%,  Thinner ice shell & 0.2  & 132   & 0.02  & 0.09 & 35   & 0.02    & 0.1     \\ 
MgSO$_4$ 10 Wt\%, Thicker ice shell & 0.6  & 18    & 0.002 & 0.01 & 35   & $\ll 1$ & $\ll 1$ \\
MgSO$_4$ 10 Wt\%, Thinner ice shell & 1.5  & 130   & 0.02  & 0.09 & 35   & 0.2   & 0.8     \\ 
\hline
\end{tabular}
\caption{Assumed properties and resulting calculated upper bounds on the strengths of the magnetic fields induced by oceanic fluid flows. Ambient magnetic field strengths, $F_o$, from \citet{ShowmanMalhotra99}; radial and theta flow speeds, $U_r$ and $U_{\theta}$ with $U=\Omega D Ro$, from Figure~\ref{fig:oceanflows}; ocean thicknesses, $D$, from \citet{Vance_2018}; and electrical conductivity, $\sigma$, from Figure~\ref{figure:conductivity}. These signals are anticipated to be largest near the equator where $U_\theta$ and $U_r$ are strongest, as indicated in Figure~\ref{fig:oceanflows}b-c.}
\label{table:flow}
\end{table}


The simplified approach shown above gives an order of magnitude estimate of the maximum induced field. Future work will assess the implications of these assumptions through more detailed calculations. For example, we have assumed a homogeneous and constant Jovian field; however, the magnetic environment throughout the orbit close in to Jupiter may be highly variable and the external field is affected by the presence of heavy ions and a variable magnetosphere dynamics throughout a single orbit \citep[e.g.,][]{schilling2008influence}. The temporal and spatial variation of the ambient field is expected to be significant and the influence of these variations on ocean flow-driven magnetic field signatures remains to be explored. Kinematic models that directly solve the coupled momentum and induction equations are also an exciting avenue to refine these estimates.

\section{Discussion and Conclusions}\label{section:discussion}


The inverse problem of reconstructing the full induction response is complex and is discussed in detail elsewhere \citep[e.g.,][ and Cochrane et al. \textit{in progress}]{khurana2009electromagnetic}. Here, we focus instead on how the adiababic conductivity profile of the ocean affects the induction response relative to the mean case that is usually considered  in space physics analyses \citep[e.g.,][]{kivelson2000galileo}, and relative to the isothermal case often considered in analyses of interior structure \citep[e.g.,][]{schubert2004interior}.

Differences between the adiabatic and mean conductivity cases have less dependence on frequency (Tables~\ref{table:inducedEuropa}-\ref{table:inducedCallisto} and Figures~S1, S3, and S5). For Europa, the nominal oceans with ice shells 5- and 30-km thick have errors of about 6\% and 3\%, respectively, and amount to nearly a 1 nT difference for the largest signals that exceed 20~nT. For Ganymede, the nominal oceans with ice shells $\sim$25- and $\sim$100-km thick have errors of about 7\% and 3\%, and are also nearly 1~nT for the largest signals that exceed 10~nT. For Callisto, the induction response of the mean conductivity ocean for ice shells of $\sim$100- and $\sim$130-km thickness is within about 2\% of the response for the adiabatic ocean, less than 0.3~nT for the largest signals that approach 10~nT.

The induction response of the adiabatic ocean differs from that of the equivalent ocean with the conductivity of fluid at the ice-ocean interface. The greater mismatch of conductivities of the lower part of the ocean causes large differences in amplitude and phase at lower frequencies (i.e. for larger skin depths).  For Europa, this means that the lower-frequency mean-motion signal (3.2$\times 10^{-6}$~Hz; Table~\ref{table:inducedEuropa}) differs by more than 15\% for the warmer lower-salinity oceans, or about 0.1~nT. For Ganymede, the differences at the mean-motion frequency (1.62$\times 10^{-6}$~Hz; Table~\ref{table:inducedGanymede}) can approach 25\%, which amounts to 0.04~nT. For Callisto, the differences at the mean-motion frequency (6.9$\times 10^{-7}$~Hz; Table~\ref{table:inducedCallisto}) approach 20\%, which amounts to only 2~pT for the small predicted residual field based on the mean field.  By contrast, the higher-frequency diurnal signals differ by less than 5\%.

Based on the circulation models and upper bound induced magnetic field estimates described in Section~\ref{section:flows}, flow-induced fields may be a prominent component of the magnetic fields measured in the low latitudes for Europa and Ganymede. The peak flow-induced magnitude is 30-40~nT (Table~\ref{table:flow}) compared with Jovian-induced residual fields of less than 20~nT for both Europa (Table~\ref{table:inducedEuropa}) and Ganymede (Table~\ref{table:inducedGanymede}).


\subsection{Implications for future missions}
The Europa Clipper mission will conduct multiple ($>$40) flybys of Europa, and will investigate its induction response with the goal of constraining the ocean conductivity to within $\pm0.5$~S~m$^{-1}$ and ice thickness to within $\pm$2~km \citep{buffington2017evolution}. The flybys at high latitudes will allow the Europa Clipper investigation to isolate flow-induced fields from the diffusive response, and possibly to derive constraints on currents in the ocean. With independent constraints on ice thickness obtained from the Radar for Europa Assessment and Sounding: Ocean to Near-surface (REASON) and Europa Imaging System (EIS) investigations \citep{Steinbrugge_2018}, it may be possible to constrain the ocean's temperature and thus the adiabatic structure for the best-fit ocean composition inferred from compositional investigations. The analyses provided here (Figure~\ref{figure:EuropaResField} and Table \ref{table:inducedEuropa}) indicate that a sensitivity of 1.5~nT is probably insufficient to distinguish between end-member MgSO$_4$ and NaCl oceans, but might be sufficient to distinguish between order-of-magnitude differences in salinity. 

The JUpiter ICy moons Explorer (JUICE) will execute two Europa flybys and nine Callisto flybys, and will orbit Ganymede \citep{grasset2013jupiter}. The magnetic field investigation seeks to determine the induction response to better than 0.1~nT. The Europa flybys might aid the Europa Clipper investigation in constraining the composition of the ocean. At Ganymede, the magnetic field investigation will not be sufficient to discern the presence of a basal liquid layer at the ice~VI-rock interface. Although the ability to discern between ocean compositions could not be assessed owing to insufficient electrical conductivity data at high pressures, it seems likely that useful constraints could be derived based on the signal strengths at Ganymede, if laboratory-derived electrical conductivity data for relevant solutions under pressure became available. At Callisto, 0.1~nT accuracy may only allow sensing of the induction response to Jupiter's synodic field, which might be sufficient to infer the thickness and salinity of an ocean if adequate temporal coverage is obtained to confirm the phase of the response.

\acknowledgments 
Work by JPL co-authors was partially supported by the Jet Propulsion Laboratory, Caltech, and by the Icy
Worlds node of NASA's Astrobiology Institute (13-13NAI7\_2-0024). This work was partially supported by NASA's Europa Clipper project. Work by MJS was supported by the NASA Earth and Space Science Fellowship Program - Grant 80NSSC18K1236. Work by KMS was also supported by NASA Grant NNX14AR28G. 

The Matlab scripts and associated data needed to compute the results shown here are currently being archived. The data and scripts for radial structure models and diffusive induction will be placed on github (https://github.com/vancesteven/PlanetProfile) and Zenodo.

The MagIC code is publicly available at the https://magic-sph.github.io/contents.html website. All global convection model data were first published in Soderlund et al. (2019) and are available therein.

\textcopyright 2020. All rights reserved.


%
%




\end{document}